\documentclass[aps,twocolumn,nofootinbib,superscriptaddress]{revtex4-1}

\usepackage{graphicx}
\usepackage{xcolor}
\usepackage{bm}
\usepackage{amsmath}
\usepackage{amssymb}
\usepackage{hyperref}

\begin{document}
 
\title{Resolving continua of fractional excitations by spinon echo in \\ THz 2D coherent spectroscopy}

\author{Yuan Wan}
\email{yuan.wan@iphy.ac.cn}
\affiliation{Institute of Physics, Chinese Academy of Sciences, Beijing 100190, China}

\author{N. P. Armitage}
\email{npa@jhu.edu}
\affiliation{Institute for Quantum Matter \& Department of Physics and Astronomy, The Johns Hopkins University, Baltimore, MD 21218, USA}
\affiliation{Japan Society for the Promotion of Science, International Research Fellow, Institute for Solid State Physics, The University of Tokyo, Kashiwa 277-8581, Japan}

\date{\today}

\begin{abstract}

We show that the new technique of terahertz 2D coherent spectroscopy is capable of giving qualitatively new information about fractionalized spin systems.  For the prototypical example of the transverse field Ising chain, we demonstrate theoretically that, despite the broad continuum of excitations in linear response, the 2D spectrum contains sharp features that are a coherent signature of a ``spinon echo'', which gives previously inaccessible information such as the lifetime of the two-spinon excited state. The effects of disorder and finite lifetime, which are practically indistinguishable in the linear optical or neutron response, manifest in dramatically different fashion in the 2D spectra. Our results may be directly applicable to model quasi-1D transverse field Ising chain systems such as CoNb$_2$O$_6$, but the concept can be applied to fractionalized spin systems in general.

\end{abstract}

\maketitle

In the recent years wholly new classes of condensed matter systems have become of intense interest. Topological materials, quantum spin liquids, and strange metals are characterized by Berry phase effects, fractional excitations, and highly entangled ground states~\cite{balents2010spin,xiao2010berry,savary2016quantum,armitage2018weyl}. However, we can measure many of their correlations only imperfectly with existing tools.  A promising direction is nonlinear response that has been used to characterize the symmetry of semiconductors~\cite{shen1984principles} and magnets~\cite{fiebig2005second}, Berry phase in topological semimetals~\cite{sipe2000second,morimoto2016topological,moore2010confinement}, and exotic ground states and excitations in correlated systems~\cite{zhao2017global,zhao2016evidence,babujian2016probing}.

\begin{figure}[t]
\includegraphics[width=0.8\columnwidth]{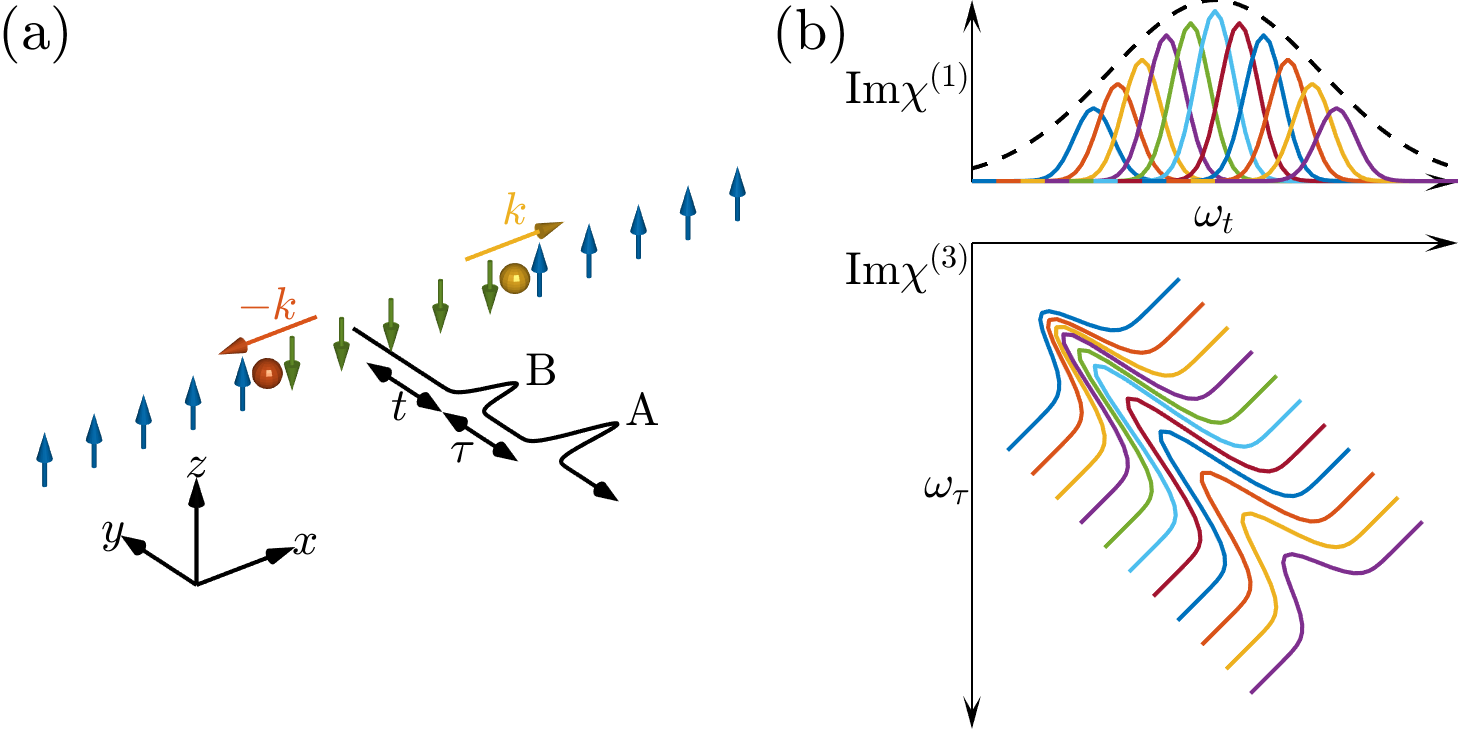}
\caption{(a) The experimental setup for THz 2D coherent spectroscopy. Two linearly polarized magnetic field pulses $A$ and $B$ arrive at the sample (in this case, transverse field Ising chain (TFIC)) at time $0$ and $\tau$. Magnetization is recorded at time $\tau + t $. In the FM phase, the pulses excite a pair of spinons (domain walls) with momenta $\pm \mathbf{k}$.  (b) Top: 1D spectroscopy probes the linear magnetic susceptibility $\chi^{(1)}(\omega)$ of TFIC. Each pair of spinons with momenta $\pm \mathbf{k}$ gives an absorption peak. The peaks congest the frequency axis, resulting in a spinon continuum. Bottom: 2D spectroscopy probes nonlinear magnetic susceptibilities of the TFIC. The signal due to the third order susceptibility $\chi^{(3)}(\omega_t, \omega_\tau)$ can resolve the spinon continuum by spreading it into the frequency plane.  Spectral congestion occurs along the diagonal, whereas the width of the individual resonance peak is revealed along the anti-diagonal direction.}
\label{fig:sketch}
\end{figure}

For quantum spin liquids, one of their most remarkable properties is the emergence of fractional particles, known as spinons, that may be understood as carrying half a conventional spin degree of freedom. Spinons present a challenge for conventional spectroscopy as they must be excited in pairs. This typically leads to a broad continuum spectrum that represent a convolution of all possible ways that energy and momentum can be shared between two spinons.  In conventional linear magnetic susceptibility  $\chi^{(1)}(\omega)$ of a spin chain~\cite{Morris14a} light excites a pair of spinons with opposite momenta (Fig.~\ref{fig:sketch}a). Each pair gives rise to a peak in the absorption spectrum $\mathrm{Im}\chi^{(1)}$ centered at the frequency $\omega = \lambda_\mathbf{k}+\lambda_{-\mathbf{k}}$, where $\lambda_\mathbf{k}$ is the dispersion relation of the spinon. As there are infinitely many such pairs, the absorption peaks congest the frequency axis, resulting in a broad continuum (Fig.~\ref{fig:sketch}b, top). While the broad continuum seen with terahertz (THz) optical spectroscopy and neutron scattering has reasonably been taken as evidence for spinons in spin chains \cite{lake2005quantum,Morris14a,Coldea2010}, the situation is less straightforward in higher dimensional spin liquid candidates e.g. 2D Kitaev materials, herbertsmithite, and triangular lattices~\cite{han2012fractionalized,banerjee2017neutron,shen2016evidence,paddison2017continuous,zhang2018hierarchy,zhu2017disorder}. In such cases, the relative importance of finite lifetime and disorder and even fractionalization itself is unclear. In all cases, the intrinsic spectral properties of spinons such as the line width and shape are hidden in the continuum.

\begin{figure*}
\includegraphics[width = 1.8\columnwidth]{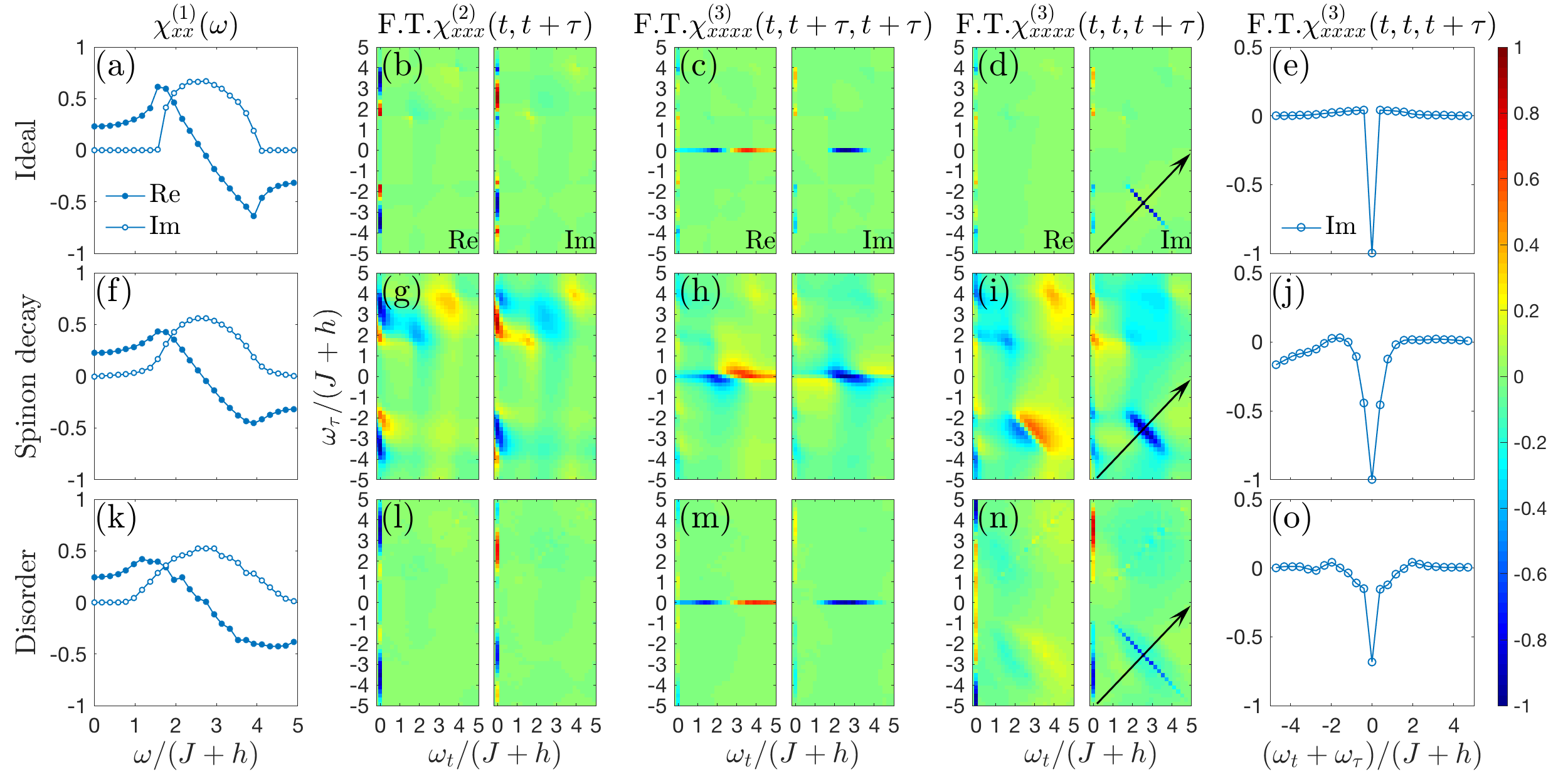}
\caption{1D and 2D spectra in the FM phase ($h/(h+J) = 0.3$) of the TFIC. From the top to bottom, the rows show the case with no dissipation ($1/T_{1,2}=0$), with dissipation ($1/T_{1,2} = 0.2(J+h)$ (other values of $T_{1,2}$ bring no significant changes), and with quenched disorder. From the left to right, the columns show respectively $\chi_{xx}^{(1)}(\omega)$, and the Fourier transforms (FT) of $\chi_{xxx}^{(2)}(t,\tau+t)$, of $\chi_{xxxx}^{(3)}(t,\tau+t,\tau+t)$, of $\chi_{xxxx}^{(3)}(t,t,\tau+t)$ and its profile along a cut indicated by the arrow. Only half of the frequency plane is shown; the other half is related by complex conjugation. For the cases without disorder (top, middle rows), the calculation is done on a chain of $L=100$ with periodic boundary condition. For the disorder case (bottom row), we set $h_n = h a_n$ and $J_n = J b_n$, where $a_n,b_n$ are site dependent, dimensionless random numbers drawn uniformly from the interval $(0.5,1.5)$. The spectra are calculated for a chain of $L=40$ with open boundaries, and averaged over 200 disorder realizations.}
\label{fig:chi_x}
\end{figure*}

In this work, we show that the new technique of THz two-dimensional coherent spectroscopy (2DCS)~\cite{kuehn2010two,woerner2013ultrafast} can provide qualitatively new information on the dynamical properties of spinons.  We explore our ideas in the context of the simplest minimal model for fractionalization -- the transverse field Ising chain (TFIC) -- but the possibilities are more general. In the optical and radio frequency range~\cite{aue1976two,cundiff2013optical, hamm2011concepts,mukamel1995principles} 2DCS is an established technique that probes nonlinear susceptibilities. Thanks to recent technical advances that enable table-top high intensity THz sources, it has been extended recently to the THz range to study graphene and quantum wells~\cite{kuehn2010two,woerner2013ultrafast}, molecular rotations~\cite{lu2016nonlinear}, and  spin waves in the conventional magnet YFeO$_3$~\cite{lu2017coherent}. THz 2DCS uses two pulses in a collinear geometry to excite a system at one frequency and detect at another, thus producing a 2D spectrum. Applications of 2DCS include quantifying nonlinear couplings between excitations and --  relevant to the present work -- separating inhomogeneous and homogeneous broadening~\cite{cundiff2013optical, hamm2011concepts, mukamel1995principles}. A similar mechanism will allow for the resolution of the spinon continuum in the 2D spectrum (Fig.~\ref{fig:sketch}b, bottom), where congestion occurs along the 2D spectrum's diagonal, but the intrinsic line width of each spinon pair is revealed by the spectral width along the anti-diagonal. 

The TFIC Hamiltonian is~\cite{Pfeuty1970}:
\begin{align}
    H = -J (\sum^{L-1}_{n=1}\sigma_n^z \sigma_{n+1}^z+\eta \sigma^z_{L} \sigma^z_1) - h\sum_n  \sigma_n^x,
    \label{eq:tfic}
\end{align}
Here $\sigma^{x,y,z}_n$ are Pauli matrices, $J>0$ is the ferromagnetic exchange,  $h>0$ is the transverse field, and $L$ is the chain length. We shall use periodic ($\eta=1$) and open ($\eta=0$) boundary conditions as they suit our purposes. Macroscopic response functions are independent of such choices. This system admits a two-fold degenerate ferromagnetic (FM) ground state for $h<J$ and a single paramagnetic (PM) ground state for $h>J$.  While strictly speaking the TFIC is not a spin liquid, the domain wall excitations of the FM phase are close analogues of spinons. Henceforth, we use ``domain wall'' and ``spinon'' interchangeably.

We consider a setup similar to that used in Ref.~\cite{lu2017coherent}. Two linearly polarized magnetic field pulses A and B arrive at the sample at time 0 and $\tau>0$ (Fig.~\ref{fig:sketch}a). The magnetization at a time $\tau + t $ along direction $\alpha$, $M^\alpha_\mathrm{AB}(\tau + t )$, is a convolution of applied field with the sample response~\cite{TimeDefinitions}. The experiment is then repeated but with pulse A or B alone and the magnetization recorded as $M^\alpha_\mathrm{A}(\tau + t )$ and $M^\alpha_\mathrm{B}(\tau + t )$. The nonlinear signal is defined as $M^\alpha_\mathrm{NL}(t,\tau) = M^\alpha_\mathrm{AB}(\tau + t )-M^\alpha_\mathrm{A}(\tau + t )-M^\alpha_\mathrm{B}(\tau + t )$. The 2D spectrum is the Fourier transform (FT) of $M_\mathrm{NL}^{\alpha}(t,\tau)$ over the domain $t>0,\tau>0$. 

The nonlinear magnetization $M_\mathrm{NL}^{\alpha}(t,\tau)$ is a direct measure of the second and/or third order magnetic susceptibilities. For simplicity, we model the  magnetic field as two Dirac-$\delta$ pulses with the  same polarization $\beta$, i.e. $B^\beta(s) = A^\beta_0 \delta(s) + A^\beta_\tau \delta(s-\tau)$, where $s$ is time, and $A^\beta_{0,\tau}$ the pulse areas. In principle, the polarizations of pulse A and B can be different. The nonlinear signal (See Supplemental Material (SM)~\cite{SupplMat}) is,
\begin{align}
    M^\alpha_\mathrm{NL}(t,\tau) &= A_0^\beta A_\tau^\beta \chi_{\alpha\beta\beta}^{(2)}(t,\tau + t ) \nonumber\\
    & + (A_0^\beta)^2 A_\tau^\beta \chi_{\alpha\beta\beta\beta}^{(3)}(t,\tau + t ,\tau + t ) \nonumber\\
    &+ A_0^\beta (A_\tau^\beta)^2 \chi_{\alpha\beta\beta\beta}^{(3)}(t,t,\tau + t )+O(A^4).
    \label{eq:signal}
\end{align} 
Here, we have retained the dominant and sub-dominant contributions. The two $\chi^{(3)}$ terms encode different physical processes.  In the first, pulse A couples to the sample at second order whereas pulse B couples at first order.  In the second, the contributions of A and B are switched. 

We are primarily interested in the spinons in the FM phase at zero temperature, and thus use the representative model parameters $h/(h+J) = 0.3$ in the ensuing discussion. Since $\sigma^{x}_n$ excites spinon pairs, we focus on the polarization $\alpha=\beta = \hat{x}$. We calculate $\chi^{(2)}_{xxx}$ and $\chi^{(3)}_{xxxx}$ analytically through the following procedure (see SM~\cite{SupplMat} for details). We map Eq.~\eqref{eq:tfic} to free, fermionic spinons by using the Jordan-Wigner transformation~\cite{Pfeuty1970}. Each pair of spinons with momenta $\pm k$ form a two-level system (TLS), whose ground (excited) state corresponds to the absence (presence) of the said pair. The energy level splitting is $2\lambda_{k}$, where $\lambda_k = 2\sqrt{J^2+h^2+2Jh\cos k}$ is the spinon dispersion. As the TLSs formed by different spinon pairs are decoupled, the TFIC is equivalent to an ensemble of independent TLSs, thereby permitting a straightforward calculation of the nonlinear response~\cite{mukamel1995principles}. 

In more realistic models, additions to Eq.~\eqref{eq:tfic} such as additional exchange interactions and spin-lattice couplings induce spinon interactions, which give effects such as spinon decay. By the TLS analogy, we incorporate these effects phenomenologically through a population time $T_1$ and decoherence time $T_2$~\cite{shen1984principles}, which captures the essential physics while maintaining the analytic tractability~\cite{SupplMat}. We assume $T_{1,2}$ are $k$-independent for simplicity. The ideal TFIC then corresponds to $T_{1,2}=0$.

We begin with the linear susceptibility per site,
\begin{align}
    \chi^{(1)}_{xx}(t) = \frac{2\theta(t)}{L}\sum_{k>0}\sin^2\theta_k e^{-t/T_2} \sin(2\lambda_k t),
    \label{eq:chi1}
\end{align}
where $\sin\theta_k = -2J\sin k/\lambda_k$ is the optical matrix element. The summation is over the positive half of the first Brillouin zone (1BZ). Using the above TLS picture, we interpret Eq.~\eqref{eq:chi1} as follows. The magnetic field pulse induces optical transitions in all TLS, producing a damped oscillatory signal with frequency $2\lambda_k$. The damping coefficient $1/T_2$ reflects the spinon decay. Since the frequency takes its value from a dense spectrum given by $k$, dephasing leads to an additional decay of $\chi_{xx}^{(1)}(t)$, which is difficult to distinguish from the intrinsic decay due to finite $T_2$.  This difficulty of unravelling the dephasing and the intrinsic decay persists in the frequency domain. Here, each spinon pair contributes an absorption peak in $\mathrm{Im}\chi^{(1)}_{xx}$ centered at the energy $2\lambda_k$ with width $1/T_2$. As $k$ runs over the 1BZ, the peaks form a broad continuum, which disguises the intrinsic line width $1/T_2$. Comparing the continuum for $1/T_2 = 0$ (Fig.~\ref{fig:chi_x}a) and $1/T_2 = 0.2(J+h)$ (Fig.~\ref{fig:chi_x}f), the difference is merely quantitative.

We then turn to the lowest order nonlinear response:
\begin{align}
 &   \chi_{xxx}^{(2)}(t,\tau + t ) = \frac{4\theta(t)\theta(\tau)}{L}\sum_{k>0}\sin^2 \theta_k \cos\theta_k \nonumber\\
&    \times[e^{-t/T_1}\cos(2\lambda_k \tau)-e^{-(t+\tau)/T_2}\cos(2\lambda_k(\tau + t ))].
    \label{eq:chi2}
\end{align}
The first term on the right hand side of Eq.~\eqref{eq:chi2} is non-oscillatory in $t$. In the frequency domain, this gives rise to a peak centered at $\omega_t = 0$, appearing as the streak along the $\omega_\tau$ axis shown in Fig.~\ref{fig:chi_x}b. Increasing $1/T_1$ from 0 leads to broadening of the streak (Fig.~\ref{fig:chi_x}g). Viewing $\omega_\tau$ as the pumping frequency and $\omega_t$ the detecting frequency, this streak is a THz rectification (TR) signal~\cite{lu2017coherent}. The second term of Eq.~\eqref{eq:chi2} is oscillatory in $t+\tau$. Yet, similar to Eq.~\eqref{eq:chi1}, the dephasing leads to decay, which is further modulated by the intrinsic decay due to $T_2$. This results in a diffusive, barely discernible signal in the first frequency quadrant (Fig.~\ref{fig:chi_x}b,g), which is similar to the non-rephasing (NR) signal usually found in $\chi^{(3)}$~\cite{lu2017coherent}. See SM~\cite{SupplMat} for detailed discussion of these features.
 
\begin{figure}
    \centering
    \includegraphics[width = 0.8\columnwidth]{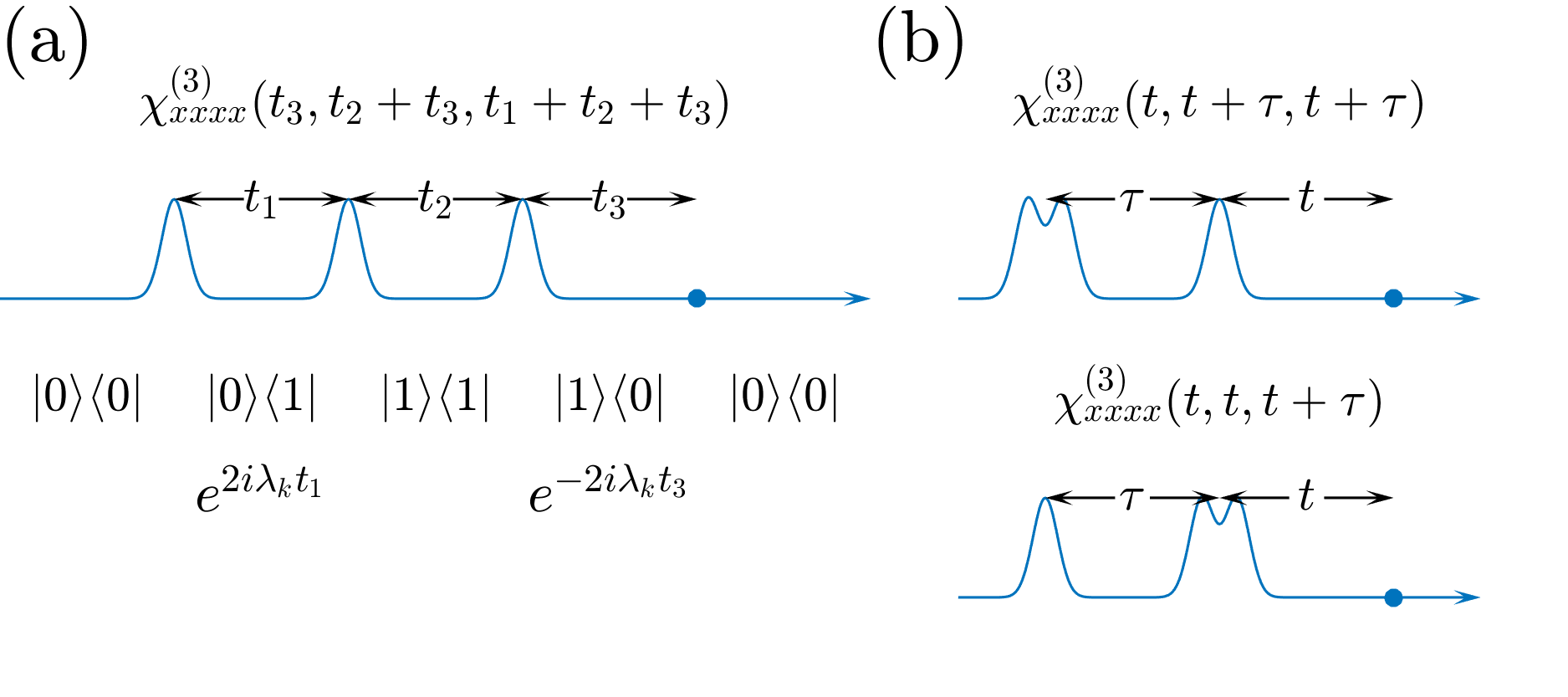}
    \caption{(a) Three pulse process associated with $\chi^{(3)}_{xxxx}(t_3,t_2+t_3,t_1+t_2+t_3)$.  The spinon echo process that produces the rephasing signal $A^{(4)}_k$ (Eq.~\eqref{eq:chi3}) is also shown. 0 (1) stands for the ground (excited) state of the two-level system formed by the spinon pair $\pm k$. The density matrices during $t_1$ and $t_3$ are Hermitian conjugate partners, and thus their time evolution are effectively time reversals of each other. (b) The $\chi^{(3)}$ terms measured in the two-pulse set up (Eq.~\eqref{eq:signal}) are special limits of the three-pulse process.}
    \label{fig:spinon_echo}
\end{figure}

Qualitatively different physics appears in $\chi^{(3)}_{xxxx}$. It is instructive to consider the more general form  that corresponds to a three-pulse process (Fig.~\ref{fig:spinon_echo}a): $\chi_{xxxx}^{(3)}(t_3,t_2+t_3,t_1+t_2+t_3) = - (\theta(t_1)\theta(t_2)\theta(t_3)/L ) \sum_{k>0}A^{(1)}_k+A^{(2)}_k+A^{(3)}_k+A^{(4)}_k$, where
\begin{subequations}
\begin{align}
    A^{(1)}_k &= 8\sin^2 \theta_k \cos^2 \theta_k \sin(2\lambda_k(t_3+t_2+t_1)) 
    \nonumber\\
    {} & \times e^{-(t_1+t_2+t_3)/T_2};
   \\
    A^{(2)}_k &= -8\sin^2 \theta_k \cos^2 \theta_k \sin(2\lambda_k(t_2+t_1))
    \nonumber\\
    {} & \times e^{-(t_1+t_2)/T_2}e^{-t_3/T_1}; 
    \\
    A^{(3)}_k &= 4\sin^4 \theta_k \sin(2\lambda_k(t_3+t_1))
    \nonumber\\
    &\times e^{-(t_1+t_3)/T_2}e^{-t_2/T_1};
    \\
    A^{(4)}_k &= 4\sin^4 \theta_k \sin(2\lambda_k(t_3-t_1))
    \nonumber\\
    &\times e^{-(t_1+t_3)/T_2}e^{-t_2/T_1}.
\end{align}
\label{eq:chi3}
\end{subequations} 

$A^{(1\sim4)}_{k}$ encode distinct evolution paths of the density matrix of the spinon pair with momenta $\pm k$ due to the THz pulses. While the forms of $A^{(1,2,3)}_{k}$ resemble that of $\chi^{(2)}_{xxx}$,  $A^{(4)}_k$ is different in that $t_1$ and $t_3$ appear with opposite signs. Regardless the oscillation frequency $2\lambda_k$, the phase accumulated between the first and the second pulses ($t_1$) is cancelled after the third pulse at $t_3 = t_1$. Said differently, the dephasing process during $t_1$ is countered by the rephasing process during $t_3$. This rephasing process is the incarnation of the photon echo in the context of spinon dynamics. Tracing $A^{(4)}_k$ back to its originating density matrix evolution sequence (Fig.~\ref{fig:spinon_echo}a), we find the sequence is identical to the photon echo process from a TLS~\cite{Kurnit1964,mukamel1995principles}. Therefore, we term this process the ``spinon echo". 

Photon echo and its analogues are a sensitive diagnostics of dissipation~\cite{Kurnit1964,mukamel1995principles}. Here, the rephasing signal from the spinon echo allows for a direct measurement of $T_2$. To see this, we return to the $\chi_{xxxx}^{(3)}$ measured in the two-pulse set up (Eq.~\eqref{eq:signal}). $\chi_{xxxx}^{(3)}(t,t,\tau+t)$ corresponds to the limit $t_1\to \tau,t_2\to0,t_3\to t$ (Fig.~\ref{fig:spinon_echo}b). We may write $\sum_k A^{(4)}_k = f(t-\tau) \exp(-(t+\tau)/T_2)$, where $f(t-\tau)$ comes from the sum of $\sin^4\theta_k \sin(2\lambda_k(t-\tau))$ and decreases as $|t-\tau|$ increases due to dephasing. Crucially, the arguments of $f$ and the $T_2$ term are \emph{orthogonal} linear combinations of $t$ and $\tau$. The FT of $f$ is a broad continuum that depends on $\omega_t-\omega_\tau$, whereas the the FT of the $T_2$ term is a narrow Lorentzian function of $\omega_t+\omega_\tau$. The product of the two thus gives rise to a streak of rephasing signal in the imaginary part of the FT of $\chi_{xxxx}^{(3)}(t,t,\tau+t)$. The streak runs along the diagonal of the fourth quadrant, mirroring the energy range of spinon pairs. The width of the streak along the anti-diagonal is a direct measure of $1/T_2$: In the limit of $T_2\to 0$, the anti-diagonal width vanishes, reflecting the perfect phase cancellation in the spinon echo (Fig.~\ref{fig:chi_x}d,e); With finite $T_2$, imperfect phase cancellation leads to a finite anti-diagonal width that scales with $1/T_2$ (Fig.~\ref{fig:chi_x}f,g).

By contrast, $\chi_{xxxx}^{(3)}(t,\tau+t,\tau+t)$, corresponding to the limit $t_1\to 0, t_2\to \tau, t_3\to t$, does not contain a spinon echo (Fig.~\ref{fig:spinon_echo}b). In the limit of $1/T_1 \to 0$, $A^{(3,4)}_k$ are functions of $t_3 = t$ only. In the frequency domain, this leads to a Dirac-$\delta$ peak on the $\omega_\tau = 0$ line, which appears in the imaginary part as a streak along the $\omega_t$ axis (Fig.~\ref{fig:chi_x}c). Taking $\omega_\tau$($\omega_t$) as the pumping (detecting) frequency, this may be interpreted as a pump-probe signal~\cite{lu2017coherent}. Increasing $1/T_1$ broadens the signal (Fig.~\ref{fig:chi_x}h). 

Both $\chi^{(3)}_{xxxx}$'s contain additional features that arise from $A^{(1,2,3)}_k$ terms in Eq.~\eqref{eq:chi3}. Their FT contain a diffusive, weak NR signal in the first quadrant. They also contain a weak TR-like signal on the $\omega_\tau$ axis, which we discuss further in SM~\cite{SupplMat}.

To recap, the rephasing signal from the spinon echo process can directly reveal the $T_2$ time of spinon pairs. Crucially, in the absence of dissipation ($1/T_{1,2}=0$), the anti-diagonal width of the rephasing signal is zero. We now show that this feature is robust against quenched disorder. To this end, we set the transverse field $h_n$ and exchange constant $J_n$ to be site dependent, namely $h_n = h a_n$, $J_n = J b_n$, where $a_{n},b_{n}$ are dimensionless random numbers drawn from a uniform distribution in the interval $[0.5,1.5]$. The linear response (Fig.~\ref{fig:chi_x}k) shows only small changes comparing to the clean case. Since this model remains integrable, the spinons are still exact eigenstates, and therefore the anti-diagonal width of the rephasing signal remains resolution limited (Fig.~\ref{fig:chi_x}n,o). Its strong sensitivity to dissipation, protected by the robustness against disorder, shows the utility of 2DCS.

In the FM phase, the $\sigma^y_n$ operators can also excite spinon pairs. We therefore expect the 2DCS spectrum with $\hat{y}$ polarization to be similar to $\hat{x}$. However, since the $\sigma^y_n$ is a non-local operator in the spinon basis, the analytic treatment made for $\hat{x}$ does not translate directly to $\hat{y}$. Nevertheless, as shown in the SM~\cite{SupplMat}, numerical calculation finds that the 2D spectra along $\hat{y}$ in the FM phase (Fig.~S4) are qualitatively similar to that of $\hat{x}$. Note that in the PM phase the streak-like rephasing signal that is characteristic of fractional excitations is absent.  $\chi^{(3)}_{yyyy}$ instead shows sharp isolated peaks~\cite{SupplMat} that are typical of nonlinear spin waves~\cite{lu2017coherent,babujian2016probing}.

Using the TFIC as a prototypical example, we have demonstrated that THz 2DCS can resolve the spinon continuum and directly reveal the intrinsic line width of spinon pairs. We expect spinon echo to be a generic 2D spectral feature of models that host spinons. Provided that the spinons are coherent quasiparticles, the TLS picture naturally extends to higher dimensional spin liquids. In general, spinon echo will produce a rephasing streak qualitatively similar to that of the TFIC with finite $T_{1,2}$. In particular, the finite anti-diagonal width of the streak reflects the imperfect phase cancellation due to finite quasiparticle lifetime.

Our results may be applicable to CoNb$_2$O$_6$, which is the best known material example of a quasi-1D FM Ising chain~\cite{Coldea2010,Morris14a,Viirok2018}.  CoNb$_2$O$_6$ orders at temperatures below $\sim3$ K, but at slightly higher temperatures, the linear response is characterized by a broad lineshape characterized by superimposed 2- and 4-spinon continua, that hide information about spinon lineshapes. We expect that THz 2DCS can reveal the intrinsic spectral properties of spinons in this system. Experiments can be done in essentially the same fashion as previous THz 2DCS measurements on the conventional magnet YFeO$_3$~\cite{lu2017coherent}.  Such experiments are underway. Analyzing theoretically the 2D spectra of more realistic material models will also prove fruitful. The spinon interactions present in these models will produce additional spectral features that are beyond the minimal model considered here.

With the information gained by establishing the technique on TFIC and its material realizations, we expect even richer information can be gained by applying THz 2DCS to higher dimensional materials that are suspected to harbor a spin liquid, but have only been characterized spectroscopically as having broad lineshapes such as 2D Kitaev magnets \cite{banerjee2017neutron}, herbertsmithite \cite{han2012fractionalized}, and triangular lattices \cite{shen2016evidence,paddison2017continuous,zhang2018hierarchy,zhu2017disorder}. By direct analogy to the present results, we expect that one can measure the intrinsic lifetime of the multispinon excitations. Sharp anti-diagonal features may give direct evidence for fractionalized excitations and may be readily distinguished from highly damped conventional spin waves that could alternatively be present.   Finally, we want to stress that our work is just an early step in understanding the utility of THz 2DCS for quantum materials.  We believe important applications will be found in many systems including superconductors and topological materials.

\begin{acknowledgments} 

We thank F. Mahmood, P. Orth, M. Oshikawa, and Y. Sizyuk for discussions. YW thanks the support of the National Supercomputer Center in Tianjin, China. The numerical calculations were performed on TianHe-1A. NPA was supported as part of the Institute for Quantum Matter, an EFRC funded by the U.S. DOE, Office of BES under DE-SC0019331.  NPA acknowledges additional support from the Japan Society for the Promotion of Science, International Research Fellows Program.

\end{acknowledgments} 

\bibliography{2DTHzIsing}

\clearpage

\appendix

\onecolumngrid

\begin{center}
\textbf{Supplemental Material for \\ ``Resolving continua of fractional excitations by spinon echo in \\ THz 2D coherent spectroscopy"}
\end{center}

\section{Expressing the 2D spectra in terms of nonlinear susceptibilities}
 
In this section, we express the 2D spectra in terms of the nonlinear susceptibilities.   In the setup considered in the main text, two linearly-polarized magnetic field pulses A and B arrive at the sample at time 0 and $\tau>0$. We assume the pulse profile takes the form of the Dirac-$\delta$ function for the sake of simplicity. The magnetic field pulse is given by:
\begin{align}
B^\beta (s) = A^\beta_0 \delta(s)+A^\beta_\tau \delta(s-\tau),
\label{eq:pulse_profile}
\end{align}
where $\beta$ labels the polarization, $s$ is time, and $A_{0,\tau}$ are the pulse areas. 

The pulse field induces a response in the magnetization of the sample. The latter is related to the former through the linear and nonlinear magnetic susceptibilities:
\begin{align}
M^\alpha_\mathrm{AB}(t+\tau) &= \int^\infty_{-\infty} ds_1 \chi^{(1)}_{\alpha\beta}(t+\tau-s_1) B^\beta(s_1) + \int^\infty_{-\infty}ds_1 \int^\infty_{-\infty}ds_2 \chi^{(2)}_{\alpha\beta\beta}(t+\tau-s_1,t+\tau-s_2) B^\beta(s_1) B^\beta(s_2)\nonumber\\
&+ \int^\infty_{-\infty}ds_1 \int^\infty_{-\infty}ds_2 \int^\infty_{-\infty}ds_3 \chi^{(3)}_{\alpha\beta\beta\beta}(t+\tau-s_1,t+\tau-s_2,t+\tau-s_3)B^\beta(s_1) B^\beta(s_2) B^\beta(s_3) + \cdots.
\end{align} 
Here we have retained the leading and sub-leading nonlinear responses. $\alpha$ labels the spatial component of the magnetization. Subscript AB signifies the fact that the response mixes the effects due to both pulses. $t+\tau$ is the time at which the sample magnetization is measured/probed. $t>0$ is the delay between the second pulse (pulse B) and the time of measurement. Note that causality implies that $\chi^{(1)}_{\alpha\beta}(t+\tau-s_1) = 0$ if $s_1>t+\tau$. Similarly, $\chi^{(2)}_{\alpha\beta}(t+\tau-s_1,t+\tau-s_2) = 0$ if $s_1>t+\tau$ or $s_2>s_1$; $\chi^{(3)}_{\alpha\beta}(t+\tau-s_1,t+\tau-s_2,t+\tau-s_3) = 0$ if $s_1>t+\tau$, or $s_2>s_1$, or $s_3>s_2$.

Substituting Eq.~\eqref{eq:pulse_profile} in and using the causal properties of the susceptibilities, we find
\begin{align}
M^\alpha_{\mathrm{AB}}(t+\tau) &= \chi^{(1)}_{\alpha\beta}(t+\tau)A^\beta_{0}+\chi^{(1)}_{\alpha\beta}(t)A^\beta_{\tau} \nonumber\\
& + \chi^{(2)}_{\alpha\beta\beta}(t+\tau,t+\tau) (A^\beta_0)^2 + \chi^{(2)}_{\alpha\beta\beta}(t,t+\tau) A^\beta_\tau A^\beta_0 + \chi^{(2)}_{\alpha\beta\beta}(t,t) (A^\beta_\tau)^2 \nonumber\\
& + \chi^{(3)}_{\alpha\beta\beta\beta}(t+\tau,t+\tau,t+\tau)(A^\beta_0)^3 + \chi^{(3)}_{\alpha\beta\beta\beta}(t,t+\tau,t+\tau) A^\beta_\tau (A^\beta_0)^2 \nonumber\\
& + \chi^{(3)}_{\alpha\beta\beta\beta}(t,t,t+\tau) (A^\beta_\tau)^2 A^\beta_0 + \chi^{(3)}_{\alpha\beta\beta}(t,t,t)(A^\beta_\tau)^3 + \cdots.
\end{align}
Setting $A^\beta_\tau\to 0$ ($A^\beta_0\to 0$) yields the response in the presence of the pulse A(B) alone:
\begin{align}
M^\alpha_{\mathrm{A}}(t+\tau) &= \chi^{(1)}_{\alpha\beta}(t+\tau)A^\beta_{0} + \chi^{(2)}_{\alpha\beta\beta}(t+\tau,t+\tau) (A^\beta_0)^2 + \chi^{(3)}_{\alpha\beta\beta\beta}(t+\tau,t+\tau,t+\tau)(A^\beta_0)^3+\cdots.\\
M^\alpha_{\mathrm{B}}(t+\tau) &= \chi^{(1)}_{\alpha\beta}(t)A^\beta_\tau + \chi^{(2)}_{\alpha\beta\beta}(t,t)(A^\beta_\tau)^2 + \chi^{(3)}_{\alpha\beta\beta\beta}(t,t,t)(A^\beta_\tau)^3+\cdots.
\end{align}

The nonlinear magnetization $M^{\alpha}_{NL}$ isolates the cross-correlation between the effects of the two pulses, which is defined as:
\begin{align}
M^\alpha_\mathrm{NL}(t+\tau) &\equiv M^\alpha_\mathrm{AB}(t+\tau) - M^\alpha_\mathrm{A}(t+\tau) - M^\alpha_\mathrm{B}(t+\tau) \nonumber\\
& = \chi^{(2)}_{\alpha\beta\beta}(t,t+\tau) A^\beta_\tau A^\beta_0 + \chi^{(3)}_{\alpha\beta\beta\beta}(t,t+\tau,t+\tau) A^\beta_\tau (A^\beta_0)^2 + \chi^{(3)}_{\alpha\beta\beta\beta}(t,t,t+\tau) (A^\beta_\tau)^2 A^\beta_0 + \cdots.
\end{align}
This is the expression given in the main text.

\section{Calculating the 2D spectra along $\hat{x}$: without disorder}

In this section, we provide the details of our calculation for the linear and nonlinear susceptibilities in the ferromagnetic phase of the quantum Ising chain without quenched disorder. In this specific case, we use periodic boundary condition for the spins. The Hamiltonian is
\begin{align}
H = -J\sum^{L-1}_{n=1} \sigma^z_n \sigma^z_{n+1}-J\sigma^z_{L}\sigma^z_1-h\sum_n \sigma^x_n.
\label{eq:tfim}
\end{align}

\subsection{Diagonalizing the Hamiltonian}

The first step is to diagonalize the Hamiltonian Eq.~\eqref{eq:tfim} following the usual treatment for the TFIC~\cite{Pfeuty1970}.  To this end, we employ the Jordan-Wigner transformation:
\begin{align}
\sigma^z_n = (c^\dagger_n+c^{\phantom{\dagger}}_n) \exp(i\pi\sum^{n-1}_{j=1}c^\dagger_jc_j);\quad
\sigma^x_n = 2c^\dagger_n c^{\phantom{\dagger}}_n-1,
\end{align}
where $c^{\phantom{\dagger}}_n$ and $c^\dagger_n$ are the usual fermion annihilation/creation operators. The Hamiltonian reads:
\begin{align}
H = -J\sum_{n}(c^\dagger_n-c^{\phantom{\dagger}}_n)(c^\dagger_{n+1}+c^{\phantom{\dagger}}_{n+1})-h\sum_{n}(c^\dagger_n c^{\phantom{\dagger}}_n - c^{\phantom{\dagger}}_n c^\dagger_n).
\end{align}
The $Z_2$ symmetry of the transverse field Ising model now becomes the fermion parity symmetry. It is known that the ground state lies in the parity even sector. This implies that the Jordan-Wigner fermions must obey anti-periodic boundary condition.

We switch to the momentum representation, $c_n = (1/\sqrt{L})\sum_{k}c_k \exp(ikn)$. The momentum $k = (2m+1)\pi/L$, $m\in Z$. Substituting in one gets, 
\begin{align}
H = \sum_{k>0}(c^\dagger_k,c_{-k})\begin{pmatrix}
\epsilon_k & i\Delta_k \\
-i\Delta_k & -\epsilon_k
\end{pmatrix}
\begin{pmatrix}
c_k\\
c^\dagger_{-k}
\end{pmatrix},
\end{align}
where $\epsilon_k = -2J\cos k-2h$, $\Delta_k = -2J\sin k$. The summation is restricted to the positive half of the first Brillouin zone. We perform the Bogoliubov transformation:
\begin{align}
\begin{pmatrix}
c_k\\
c^\dagger_{-k}
\end{pmatrix} = 
\begin{pmatrix}
\displaystyle \cos\frac{\theta_k}{2} & \displaystyle -i\sin\frac{\theta_k}{2} \\
\displaystyle -i\sin\frac{\theta_k}{2} & \displaystyle \cos\frac{\theta_k}{2}
\end{pmatrix}
\begin{pmatrix}
d_k\\
d^\dagger_{-k}
\end{pmatrix},
\end{align}
where $\cos\theta_k = \epsilon_k/\lambda_k$, $\sin\theta_k = \Delta_k/\lambda_k$, $\lambda_k = \sqrt{\epsilon^2_k+\Delta^2_k}$. The Hamiltonian is now diagonalized in the spinon or $d_k$ basis:
\begin{align}
H = \sum_{k>0}\lambda_k (d^\dagger_k d^{\phantom{\dagger}}_{k}-d^{\phantom{\dagger}}_{-k}d^\dagger_{-k}) = \sum_{k}\lambda_k (d^\dagger_k d_k-\frac{1}{2}).
\end{align}
The ground state is annihilated by all $d_k$. 

For later purposes, we may also express $M^x$, the total magnetization along $x$ direction, in the spinon basis:
\begin{align}
M^x = \frac{1}{2}\sum_{n}\sigma^x_n = \sum_{k>0}m^x_k,
\end{align}
where
\begin{align} 
m^x_k &\equiv c^\dagger_k c_k - c_{-k}c^\dagger_{-k}
= (c^\dagger_k, c_{-k})\begin{pmatrix}
1 & 0\\
0 & -1
\end{pmatrix}\begin{pmatrix}
c_k \\
c^\dagger_{-k}
\end{pmatrix} 
\nonumber\\
& = (d^\dagger_k,d_{-k})
\begin{pmatrix}
\displaystyle \cos\frac{\theta_k}{2} & \displaystyle i\sin\frac{\theta_k}{2} \\
\displaystyle i\sin\frac{\theta_k}{2} & \displaystyle \cos\frac{\theta_k}{2}
\end{pmatrix}
\begin{pmatrix}
1 & 0\\
0 & -1
\end{pmatrix} 
\begin{pmatrix}
\displaystyle \cos\frac{\theta_k}{2} & \displaystyle -i\sin\frac{\theta_k}{2} \\
\displaystyle -i\sin\frac{\theta_k}{2} & \displaystyle \cos\frac{\theta_k}{2}
\end{pmatrix}
\begin{pmatrix}
d_k \\
d^\dagger_{-k}
\end{pmatrix}
\nonumber\\
& = (d^\dagger_k,d_{-k})
\begin{pmatrix}
\cos\theta_k & -i\sin\theta_k \\
i\sin\theta_k & -\cos\theta_k
\end{pmatrix}
\begin{pmatrix}
d_k \\
d^\dagger_{-k}
\end{pmatrix}.
\end{align}

It is instructive to use the Anderson pseudo-spins: 
\begin{align}
\tau^x_k \equiv d_{-k}d_k+d^\dagger_{k}d^\dagger_{-k},
\quad 
\tau^y_k \equiv i(d_{-k}d_k-d^\dagger_{k}d^\dagger_{-k}),
\quad
\tau^z_k \equiv d^\dagger_k d_k - d_{-k}d^\dagger_{-k}.
\end{align}
The two-dimensional Hilbert space is spanned by the absence (pseudo-spin $\tau^z_k=-1$) and presence (pseudo-spin $\tau^z_k = 1$) of the spinon pair with momenta $k,-k$. Using the Anderson pseudo-spins, the Hamiltonian reads:
\begin{align}
H = \sum_{k>0} \lambda_k \tau^z_k,
\end{align}
which describes an ensemble of independent two-level systems. The energy level splitting is $2\lambda_k$, which corresponds to the energy cost for exciting a pair of spinons out of vacuum. The $m^{x}_k$ operator reads:
\begin{align}
m^x_k = \cos\theta_k \tau^z_k+\sin\theta_k \tau^y_k.
\end{align}
In particular, $m^x_k$ contains a term that switches the ground state  and the excited state. In the Heisenberg picture:
\begin{align}
m^x_k(t) = \cos\theta_k \tau^z_k + \sin\theta_k (\tau^y_k\cos2\lambda_k t + \tau^x_k \sin2\lambda_k t).
\end{align}

\subsection{Computing the susceptibilities\label{sec:kubo}}

We are now ready to calculate the various linear and nonlinear susceptibilities along $\hat{x}$. For now, we consider the standard transverse field Ising chain without any dissipation. Therefore, the spinons, being exact eigenstates of the Hamiltonian, possess infinite lifetime.

We first consider the linear susceptibility $\chi^{(1)}_{xx}(t)$. The starting point is the Kubo formula:
\begin{align}
\chi^{(1)}_{xx}(t) &= \frac{i\theta(t)}{L}\langle [M^x(t),M^x(0)] \rangle = \frac{i\theta(t)}{L}\sum_{k>0} \langle [m^x_k(t),m^x_k(0)] \rangle \nonumber\\
& = \frac{2\theta(t)}{L}\sum_{k>0} \sin\theta_k \cos\theta_k (1-\cos(2\lambda_kt))\langle \tau^x_k\rangle + \sin\theta_k\cos\theta_k \sin(2\lambda_k t)\langle \tau^y_k \rangle - \sin^2 \theta_k \sin(2\lambda_kt)\langle \tau^z_k\rangle \nonumber\\
& = \frac{2\theta(t)}{L}\sum_{k>0}\sin^2 \theta_k \sin(2\lambda_kt).
\end{align}
Here, $\langle\cdots\rangle$ stands for the average in the ground state. The second equality follows from the fact that $m^x_k$ with different $k$ commute; the third equality follows from the Pauli algebra; the last equality follows from the property of the ground state that $\langle \tau^x_k\rangle = \langle \tau^y_k\rangle = 0$, and $\langle\tau^z_k\rangle = -1$.

The second and third order nonlinear susceptibilities can be found in the same vein. The Kubo formula for the second order nonlinear susceptibility is given by,
\begin{align}
\chi^{(2)}_{xxx}(t,\tau+t) &= \frac{i^2\theta(t)\theta(\tau)}{L}\langle[[M^x(\tau+t),M^x(\tau)],M^x(0)]\rangle 
\nonumber\\
& = \frac{i^2\theta(t)\theta(\tau)}{L}\sum_{k>0}\langle [[m^x_k(\tau+t),m^x_k(\tau)],m^x_k(0)]\rangle.
\end{align}
The nested commutators are computed step by step following the Pauli algebra:
\begin{subequations}
\begin{align}
[m^x_k(t+\tau), m^x_k(\tau)] &= 2i\sin\theta_k\cos\theta_k[\cos(2\lambda_k(t+\tau))-\cos(2\lambda_k \tau)]\tau^x_k  \nonumber\\
& - 2i\sin\theta_k\cos\theta_k[\sin(2\lambda_k(t+\tau))-\sin(2\lambda_k \tau)]\tau^y_k
\nonumber\\
& + 2i\sin^2\theta_k\sin(2\lambda_k t)\tau^z_k.
\\
[[m^x_k(t+\tau), m^x_k(\tau)],m^x_k(0)] & = 4\{\sin\theta_k\cos^2\theta_k[\sin(2\lambda_k(t+\tau))-\sin(2\lambda_k\tau)]+\sin^3\theta_k\sin(2\lambda_k t)\}\tau^x_k 
\nonumber\\
& + 4\sin\theta_k \cos^2\theta_k [\cos(2\lambda_k(t+\tau))-\cos(2\lambda_k\tau)]\tau^y_k 
\nonumber\\
& - 4\sin^2\theta_k\cos\theta_k[\cos(2\lambda_k(t+\tau))-\cos(2\lambda_k\tau)]\tau^z_k.
\end{align}
\end{subequations}
Since $\langle\tau^x_k\rangle=\langle\tau^y_k\rangle=0$, and $\langle\tau^z_k\rangle = -1$, we find the ground state expectation value of the nested commutators:
\begin{align}
\langle [[m^x_k(t+\tau), m^x_k(\tau)],m^x_k(0)] \rangle  =  4\sin^2\theta_k\cos\theta_k[\cos(2\lambda_k(t+\tau))-\cos(2\lambda_k\tau)].
\end{align}
Substituting the above back to the Kubo formula, we obtain:
\begin{align}
\chi^{(2)}_{xxx}(t,\tau+t) = -\frac{4\theta(t)\theta(\tau)}{L}\sum_{k>0}\sin^2\theta_k \cos\theta_k [\cos(2\lambda_k (\tau+t))-\cos(2\lambda_k\tau)].
\end{align}

The Kubo formula for the third order nonlinear susceptibility is given by:
\begin{align}
\chi^{(3)}_{xxxx}(t_3,t_2+t_3,t_1+t_2+t_3) &= \frac{i^3\theta(t_1)\theta(t_2)\theta(t_3)}{L}\langle[[[M^x(t_1+t_2+t_3),M^x(t_1+t_2)],M^x(t_1)],M^x(0)]\rangle \nonumber\\
& = \frac{i^3\theta(t_1)\theta(t_2)\theta(t_3)}{L} \sum_{k>0} \langle [[[m^x_k (t_1+t_2+t_3),m^x_k (t_1+t_2)],m^x_k (t_1)],m^x_k(0)]  \rangle 
\end{align}
Similar to the calculation for $\chi^{(2)}_{xxx}$, we compute the nested commutators step by step as:
\begin{subequations}
\begin{align}
[m^x_k(t_1+t_2+t_3),m^x_k(t_1+t_2)] &= 2i\sin\theta_k\cos\theta_k[\cos(2\lambda_k(t_1+t_2+t_3))-\cos(2\lambda_k(t_1+t_2))]\tau^x_k  \nonumber\\
& - 2i\sin\theta_k\cos\theta_k[\sin(2\lambda_k(t_1+t_2+t_3))-\sin(2\lambda_k (t_1+t_2))]\tau^y_k
\nonumber\\
& + 2i\sin^2\theta_k\sin(2\lambda_k t_3)\tau^z_k.
\end{align}
Continuing on:
\begin{align}
& \phantom{=} [[m^x_k(t_1+t_2+t_3),m^x_k(t_1+t_2)],m^x_k(t_1)]
\nonumber\\
& = \Big\{ 4\sin\theta_k\cos^2\theta_k [\sin(2\lambda_k(t_1+t_2+t_3))-\sin(2\lambda_k(t_1+t_2))]+4\sin^3\theta_k\sin(2\lambda_k t_3)\cos(2\lambda_kt_1) \Big\}\tau^x_k 
\nonumber\\
& + \Big\{ 4\sin\theta_k\cos^2\theta_k[\cos(2\lambda_k(t_1+t_2+t_3))-\cos(2\lambda_k(t_1+t_2))] - 4\sin^3\theta_k\sin(2\lambda_k t_3)\sin(2\lambda_k t_1) \Big\} \tau^y_k
\nonumber\\
& - 4\sin^2\theta_k\cos\theta_k \Big \{[\cos(2\lambda_k(t_1+t_2+t_3))-\cos(2\lambda_k(t_1+t_2))]\cos(2\lambda_kt_1)
\nonumber\\
& + [\sin(2\lambda_k(t_1+t_2+t_3))-\sin(2\lambda_k(t_1+t_2))]\sin(2\lambda_kt_1) \Big \}\tau^z_k.
\end{align}
Finally,
\begin{align}
& \phantom{=} [[[m^x_k(t_1+t_2+t_3),m^x_k(t_1+t_2)],m^x_k(t_1)],m^x_k(0)] 
\nonumber\\
& = 8i \Big\{ \sin^2\theta_k\cos^2\theta_k[\sin(2\lambda_k(t_1+t_2+t_3))-\sin(2\lambda_k(t_1+t_2))] + \sin^4\theta_k \sin(2\lambda_k t_3)\cos(2\lambda_k t_1) \Big\}\tau^z_k + \cdots.
\end{align}
\end{subequations}
Here, we have omitted terms that are proportional to $\tau^x_k$ and $\tau^y_k$. These terms won't contribute the ground state average, and, therefore, won't appear in the final expression for the susceptibility. Using $\langle \tau^x_k\rangle = \langle \tau^y_k\rangle = 0$, and $\langle \tau^z_k\rangle = -1$, we find:
\begin{align}
& \phantom{=} \langle [[[m^x_k(t_1+t_2+t_3),m^x_k(t_1+t_2)],m^x_k(t_1)],m^x_k(0)] \rangle
\nonumber\\
& = -8i\sin^2\theta_k\cos^2\theta_k[\sin(2\lambda_k(t_1+t_2+t_3))-\sin(2\lambda_k(t_1+t_2))] -8i\sin^4\theta_k \sin(2\lambda_k t_3)\cos(2\lambda_k t_1).
\end{align}
Substituting the above into the Kubo formula, we ultimately obtain:
\begin{subequations}
\begin{align}
\chi^{(3)}_{xxxx}(t_3,t_2+t_3,t_1+t_2+t_3) = \frac{\theta(t_1)\theta(t_2)\theta(t_3)}{L}\sum_k A^{(1)}_{k}+A^{(2)}_{k}+A^{(3)}_{k}+A^{(4)}_{k},
\end{align}
where
\begin{align}
A^{(1)}_k &= -8\sin^2 \theta_k \cos^2 \theta_k \sin(2\lambda_k (t_3+t_2+t_1));
\\
A^{(2)}_k &= 8 \sin^2 \theta_k \cos^2 \theta_k \sin(2\lambda_k (t_2+t_1));
\\
A^{(3)}_k &= -4 \sin^4 \theta_k \sin(2\lambda_q (t_3+t_1));
\\
A^{(4)}_k &= -4 \sin^4 \theta_k \sin(2\lambda_k(t_3-t_1)).
\end{align}
\end{subequations}

\subsection{Incorporating dissipation effects\label{sec:dissipation}}

Here, we calculate the linear and nonlinear susceptibilities in the presence of spinon decay. Instead of starting from the first principle, we incorporate these effects phenomenologically, which is most conveniently achieved by using an approach akin to the optical Bloch equations~\cite{shen1984principles}. As we shall see, this phenomenological approach captures the essential features of dissipation and maintains the analytical tractability of the transverse field Ising chain. 

To see how the optical Bloch equations naturally arise in this context, we note that each and every pair of spinons with momenta $\pm k$ effectively form a two-level system. Each two-level system is dynamically decoupled from the others. The state of the spinon pair $\pm k$ is specified by the $2\times 2$ density matrix $\rho_k$,
\begin{align}
\rho_k = \begin{pmatrix}
\rho_{00} & \rho_{01} \\
\rho_{10} & \rho_{11}
\end{pmatrix},
\end{align}
where $0$ and $1$ represent respectively the ground state and the excited state of the two-level system, or in terms of spinons, the absence and presence of the spinon pair. The time evolution of $\rho_{k}$ is canonically described by the optical Bloch equation (without the driving term). Its solution gives the explicit time evolution, 
\begin{align}
\rho_{00}(t) = \rho_{00}(0)+(1-e^{-t/T_1})\rho_{11}(0),\quad
\rho_{11}(t) = e^{-t/T_1}\rho_{11}(0),\nonumber\\
\rho_{10}(t) = e^{-(1/T_2+2i\lambda_k)t}\rho_{10}(0),\quad
\rho_{01}(t) = e^{-(1/T_2-2i\lambda_k)t}\rho_{01}(0).
\end{align}
Here, $2\lambda_k$ is the energy level splitting, or the energy cost for exciting the spinon pair $\pm k$ out of vacuum. $T_1$ is the time scale of spontaneous decay of the excited state, whereas $T_2$ is the decoherence time scale. The above may be rewritten as  $\rho_k(t) = \mathcal{G}_k(t) \rho_k(0)$, where the super-operator $\mathcal{G}_k(t)$ propagates the density matrix. For the sake of simplicity, we assume $T_1$ and $T_2$ times are $k$-independent.

In what follows, we calculate the linear and nonlinear susceptibilities by taking an approach that is commonly used in the analysis of 2D coherent spectroscopy~\cite{mukamel1995principles}. Along the way, we will introduce the relevant terminology and tools.

We begin with the linear susceptibility $\chi^{(1)}_{xx}$. We expand the commutator that appear in the Kubo formula and obtain,
\begin{subequations}
\begin{align}
\chi^{(1)}_{xx}(t) = -\frac{2\theta(t)}{L}\sum_{k>0}\mathrm{Im}\langle m^x_k(t)m_k(0) \rangle = -\frac{2\theta(t)}{L}\sum_{k>0} \mathrm{Im}\mathrm{Tr}(m^x_k\rho_k(t)),
\end{align}
where
\begin{align}
\rho_k(t) = \mathcal{G}_k(t)\mathcal{L}(m^x_k)\rho_k(0).
\end{align}
\end{subequations}
Here, we take the following perspective: the correlation functions on the right hand side of the first equality represents a sequence of evolution for the density matrix $\rho_k$, known as the Liouville pathway in literature~\cite{mukamel1995principles}. $\mathcal{L}(m^x_k)$ is the super-operator that means acting $m^x_k$ on $\rho_k$ from the left. Later, we shall use the super-operator $\mathcal{R}(m^x_k)$, which means acting $m^x_k$ on $\rho_k$ from the right. $\mathcal{G}_k(t)$ is the aforementioned propagator of density matrix. $\rho_k(t)$ may be readily evaluated by using the definition of $m^x_k$ and $\mathcal{G}_k(t)$:
\begin{align}
\rho_k(t) = \mathcal{G}_k(t)\mathcal{L}(m^x_k)|0\rangle\langle0| = \sin\theta_k \mathcal{G}_k(t)|1\rangle\langle0| = e^{-(1/T_2+2i\lambda_k)t}|1\rangle\langle0|.
\end{align}
Substituting it back to the expression for $\chi^{(1)}_{xx}$, we find,
\begin{align}
\chi^{(1)}_{xx}(t) = \frac{2\theta(t)}{L}\sum_{k>0}\sin^2 \theta_k e^{-t/T_2}\sin(2\lambda_k t)\label{eq:chi1_x},
\end{align}
which is the result given in the main text.

\begin{figure}
\includegraphics[width = 0.6\textwidth]{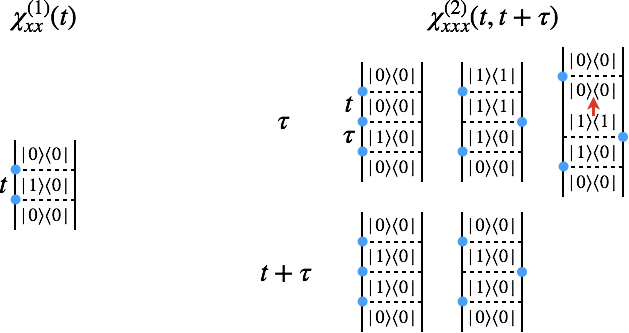}
\caption{Left: ladder diagram for the linear susceptibility $\chi^{(1)}_{xx}(t)$. Right: ladder diagrams for the second-order nonlinear susceptibility $\chi^{(2)}_{xxx}(t,\tau+t)$. The diagrams are organized in terms of their quantum phase factors. The diagrams in the top row contain phase factors that are functions of $\tau$, whereas the diagrams in the bottom row contain phase factors that are functions of $\tau+t$.}
\label{fig:chi12_diagrams}
\end{figure}

It is convenient to use a bookkeeping device that graphically represents the various Liouville pathways that contribute to $\chi^{(1)}_{xx}$. Ladder diagrams, or so-called \textit{double-sided Feynman diagrams} used commonly in 2D spectroscopy~\cite{mukamel1995principles,TomakoffNotes}, fulfills this task. Here, we have slightly tailored them to suit our purposes.  The left panel of Fig.~\ref{fig:chi12_diagrams} shows the ladder diagram that contributes $\chi^{(1)}_{xx}(t)$. We read the diagram as follows.  Time flows upward. Each rung of the ladder (dashed line) represents a transition due to the action of $m_k$. The dot on the left corresponds to the action of $\mathcal{L}(m^x_k)$, whereas the dot on the right corresponds to the action of $\mathcal{R}(m^x_k)$. Note, however, that the last dot doesn't correspond to any real transitions; it describes the final measurement process ($\mathrm{Tr}(m^x_k\rho_k)$). Representing it as a ``transition'' is mere convention. The space between two rungs is the evolution of $\rho_k$ under $\mathcal{G}_k(t)$.

For the specific diagram for linear response given in the left panel of Fig.~\ref{fig:chi12_diagrams}, the bottom of the ladder diagram is $|0\rangle\langle 0|$, corresponding to the pure ground state. Going upward, there is a transition to $|1\rangle \langle0 |$ due to the action of $m_k$. This picks up a matrix element $\langle 1|m^x_k|0\rangle$. This is followed by the evolution due to $\mathcal{G}_k(t)$, which picks up a time-evolution factor $\exp(-(1/T_2+2i\lambda_k) t)$. Finally, the measurement yields another matrix element $\langle 0|m^x_k|1\rangle$. Collecting these, the total outcome is $|\langle 1|m^x_k|0\rangle|^2\exp(-(1/T_2+2i\lambda_k) t)$, which is the term appearing in the summation of Eq.~\eqref{eq:chi1_x}.  It is important to bear in mind that other diagrams can in principle be drawn for $\chi^{(1)}_{xx}$. However, their contributions vanish due to cancellation.

We now consider the second-order nonlinear susceptibility $\chi^{(2)}_{xxx}$. A similar procedure allows us to relate the Kubo formula for $\chi^{(2)}_{xxx}(t,\tau+t)$ to the Liouville pathways. Out of a total of 9 possible pathways and their Hermitian conjugates, 5 diagrams have finite contribution to $\chi^{(2)}_{xxx}$, which are shown as ladder diagrams in the right panel of Fig.~\ref{fig:chi12_diagrams}. Comparing to the diagram for $\chi^{(1)}_{xx}$, two new elements appear: First, the spontaneous decay of the excited state $|1\rangle\langle 1|$ to the ground state $|0\rangle\langle0|$ is graphically represented as a red arrow. Second, the diagrams with \emph{odd} number of dots on the right leg have a \emph{negative} sign.

The diagrams in the right panel of Fig.~\ref{fig:chi12_diagrams} are organized in terms of their phase factors. The top row all contain a phase factor $\exp(2i\lambda_k \tau)$. The sum of these is:
\begin{subequations}
\begin{align}
& \phantom{=} [|\langle1|m^x_k|0\rangle|^2 \langle0|m_k|0\rangle - |\langle1|m^x_k|0\rangle|^2 \langle1|m^x_k|1\rangle e^{-t/T_1} - |\langle1|m^x_k|0\rangle|^2 \langle0|m^x_k|0\rangle(1-e^{-t/T_1})]e^{-\tau/T_2}\cos(2\lambda_k \tau) \nonumber\\
& = -2\cos\theta_k \sin^2 \theta_k e^{-\tau/T_2}\cos(2\lambda_k \tau) e^{-t/T_1}.
\end{align}
The bottom row all contain a phase factor $\exp(2i\lambda_k (\tau+t))$. The sum of these is:
\begin{align}
& \phantom{=} (|\langle1|m^x_k|0\rangle|^2\langle1|m^x_k|1\rangle-|\langle1|m^x_k|0\rangle|^2\langle0|m^x_k|0\rangle)\cos(2\lambda_k(\tau+t))e^{-(\tau+t)/T_2} \nonumber\\
& = 2\cos\theta_k \sin^2 \theta_k e^{-(\tau+t)/T_2}\cos(2\lambda_k(\tau+t)).
\end{align}
Combining them yields,
\begin{align}
\chi^{(2)}_{xxx}(t,\tau+t) = -\frac{4\theta(t)\theta(\tau)}{L}\sum_{k>0}\cos2\theta_k \sin^2 2\theta_k [e^{-(\tau+t)/T_2}\cos(2\lambda_k(\tau+t))-e^{-t/T_1}e^{-\tau/T_2}\cos(2\lambda_k \tau)].
\end{align}
\end{subequations}
Note the additional minus sign from the pre-factor of the Kubo formula. This is the result given in the main text. 

\begin{figure}
\includegraphics[width = 0.6\textwidth]{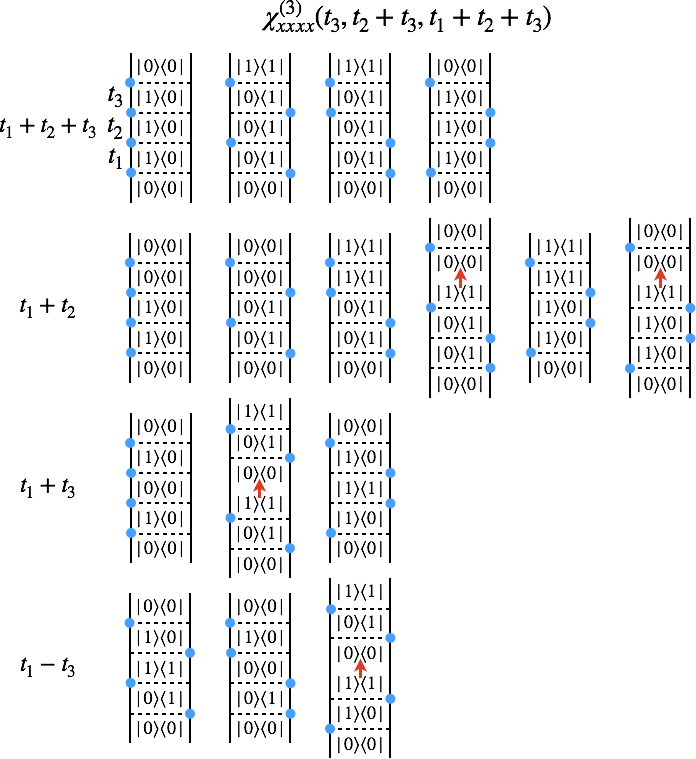}
\caption{Ladder diagrams for the third-order nonlinear susceptibility $\chi^{(3)}_{xxxx}(t_3,t_2+t_3,t_1+t_2+t_3)$. The diagrams are organized in terms of their quantum phase factors. From the top to bottom, the first, second, third, and fourth row respectively contain phase factors that are functions of $t_1+t_2+t_3$, $t_1+t_2$, $t_1+t_3$, and $t_1-t_3$.}
\label{fig:chi3_diagrams}
\end{figure}

We finally turn to the third-order nonlinear susceptibility $\chi^{(3)}_{xxxx}(t_3,t_2+t_3,t_1+t_2+t_3)$. As the complexity of the diagrams grow exponentially with the order of nonlinearity, we now have in total 42 Liouville pathways plus their Hermitian conjugates, out of which 16 independent pathways contribute. These are given in Fig.~\ref{fig:chi3_diagrams}. The diagrams are organized in terms of their phase factors: from the top to the bottom, the phase factors of the diagrams in the first, second, third, and the last row are functions of $t_1+t_2$, $t_1+t_2+t_3$, $t_1+t_3$, and $t_1-t_3$ respectively. By now all rules of the ladder diagrammatic have been given. We now employ these rules to obtain the analytic results. The sum of the first row is given by:
\begin{subequations}
\begin{align}
A^{(1)}_k &= 2[-|\langle1|m^x_k|0\rangle|^2 \langle 1|m^x_k|1\rangle^2+2|\langle 0|m^x_k|1\rangle|^2\langle 0|m^x_k|0\rangle \langle1|m^x_k|1\rangle-|\langle1|m^x_k|0\rangle|^2 \langle 0|m^x_k|0\rangle^2] \nonumber\\
& \times \sin[2\lambda_k(t_1+t_2+t_3)]e^{-(t_1+t_2+t_3)/T_2} \nonumber\\
& = -8\sin^2\theta_k \cos^2\theta_k\sin[2\lambda_k(t_1+t_2+t_3)]e^{-(t_1+t_2+t_3)/T_2}.
\end{align}
Note the factor of 2 is due to the contribution from both a diagram and its complex conjugate. The second row is:
\begin{align}
A^{(2)}_k &=2 [-|\langle1|m^x_k|0\rangle|^2\langle0|m^x_k|0\rangle\langle1|m^x_k|1\rangle + |\langle1|m^x_k|0\rangle|^2\langle0|m^x_k|0\rangle^2 + |\langle1|m^x_k|0\rangle|^2\langle1|m^x_k|1\rangle^2 e^{-t_3/T_1} \nonumber\\
&+|\langle0|m^x_k|1\rangle|^2\langle0|m^x_k|0\rangle\langle1|m^x_k|1\rangle (1-e^{-t_3/T_1})-|\langle0|m^x_k|1\rangle|^2\langle0|m^x_k|0\rangle\langle1|m^x_k|1\rangle e^{-t_3/T_1} - |\langle0|m^x_k|1\rangle|^2\langle0|m^x_k|0\rangle^2(1-e^{-t_3/T_1})]\nonumber\\
&\times\sin[2\lambda_k(t_1+t_2)]e^{-(t_1+t_2)/T_2} \nonumber\\
& = 8\sin^2\theta_k \cos^2\theta_k \sin[2\lambda_k(t_1+t_2)]e^{-(t_1+t_2)/T_2}e^{-t_3/T_1}.
\end{align}
The third row is:
\begin{align}
A^{(3)}_k &= 2[-|\langle1|m^x_k0\rangle|^4+|\langle1|m^x_k0\rangle|^4(1-e^{-t_2/T})-|\langle1|m^x_k0\rangle|^4 e^{-t_2/T_1}] \sin[2\lambda_k(t_1+t_3)]e^{-(t_1+t_3)/T_2} \nonumber\\
& = -4\sin^4\theta_k \sin[2\lambda_k(t_1+t_3)]e^{-(t_1+t_3)/T_2} e^{-t_2/T_1}.
\end{align}
The final row is:
\begin{align}
A^{(4)}_k &= 2[-|\langle0|m^x_k|1\rangle|^4e^{-t_2/T_1}-|\langle0|m^x_k|1\rangle|^4+|\langle1|m^x_k|0\rangle|^4(1-e^{-t_2/T_1})] \sin[2\lambda_k(t_1-t_3)]e^{-(t_1+t_3)/T_2} \nonumber\\
&=-4\sin^4\theta_k \sin[2\lambda_k(t_1-t_3)]e^{-(t_1+t_3)/T_2}e^{-t_2/T_1}.
\end{align}
Combing the above, we obtain,
\begin{align}
\chi^{(3)}_{xxxx}(t_3,t_2+t_3,t_1+t_2+t_3) = \frac{\theta(t_1)\theta(t_2)\theta(t_3)}{L}\sum_{k>0}A^{(1)}_k+A^{(2)}_k+A^{(3)}_k+A^{(4)}_k.
\end{align}
\end{subequations}
This is the result given in the main text. 

We conclude this section by commenting that, when taking the limit of $T_{1,2}\to \infty$, the susceptibilities agree with the results given in Sec.~\ref{sec:kubo}.

\subsection{Performing the Fourier transform \label{sec:fft}}

\begin{figure}
\includegraphics[width = 0.8\textwidth]{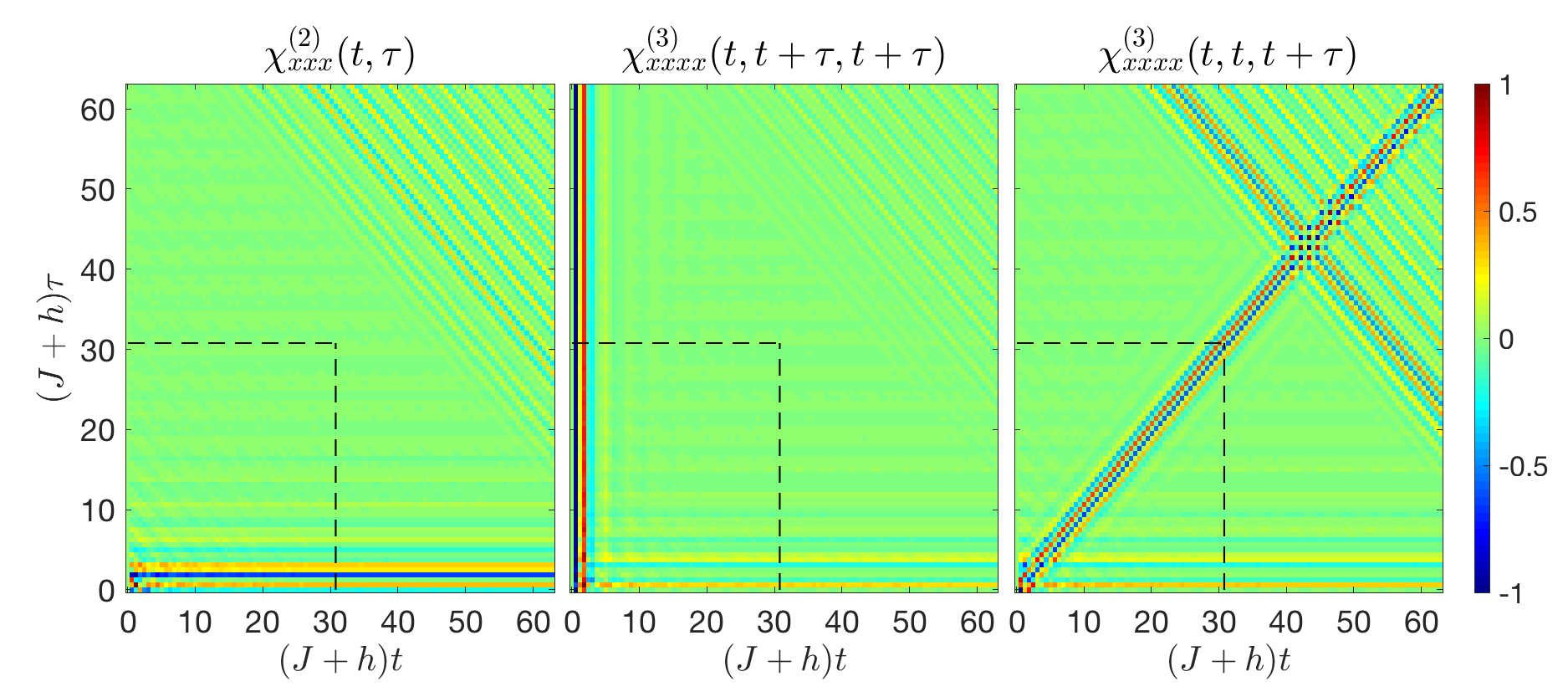}
\caption{The nonlinear susceptibilities $\chi^{(2)}_{xxx}(t,\tau)$, $\chi^{(3)}_{xxxx}(t,\tau+t,\tau+t)$, and $\chi^{(3)}_{xxxx}(t,t,\tau+t)$ as functions of $t$ and $\tau$. The revival of oscillations at late time is a finite size effect and must be excluded in the Fourier transform. The dashed box demarcates the domain over which the 2D Fourier transform is performed.}
\label{fig:real_time}
\end{figure}

With the analytic formula at hand, we are ready to compute the 2D coherent spectra. This would require performing a two-dimensional, one-sided Fourier transform with respect to the variables $t,\tau$ over the domain $t>0,\tau>0$. While it is possible to find the explicit expression directly, the resulted expressions are too complicated to be illuminating. We therefore choose a different route: we first calculate $\chi^{(2)}_{xxx}$ and $\chi^{(3)}_{xxxx}$ in the time domain using the aforementioned analytic expressions. We then perform a numerical fast Fourier transform (FFT) on the computed signal, thereby obtaining the results shown in the main text.

A couple of considerations come into choosing the time grid. Let $dt$ be the step width of the time grid. The maximal frequency that can be sampled by this grid is $\omega_\mathrm{max} = \pi/dt$. On the other hand, the maximal spinon pair energy $2\lambda_k$ is $4(J+h)$, and there is no signal above this frequency. In practice, we set $dt = \pi/(5(J+h))$, which corresponds to $\omega_\mathrm{max} = 5(J+h)$ slightly above the maximal spinon pair energy.

The maximal number of steps of the grid, on the other hand, is limited by the system size $L$. The frequency spacing $\sim O(1/L)$, which sets a time scale $\sim O(L)$. Simulating over a time longer than this time scale will result in spurious peaks in the frequency domain due to the finite system size. In practice, we determine the maximal time steps by monitoring the real time signal. The finite size effect manifests itself as a ``revival'' of oscillations at late time. We limit the FFT to times before the onset of such a revival.

Fig.~\ref{fig:real_time} shows the various nonlinear susceptibilities in real time. Here, $L=100$, $h/(J+h) = 0.3$, and $1/T_1 = 1/T_2 = 0$. The diagonal feature in $\chi^{(3)}_{xxxx}(t,t,\tau+t)$ is the rephasing signal described in the main text. The ``wavefronts'' that propagate along the diagonal at the top right corner are due to the finite size effect. We therefore perform the 2D FFT over a smaller domain demarcated by the dashed box.

\section{Calculating the 2D spectra along $\hat{x}$: with quenched disorder \label{sec:chi_x_disorder}}

Here we provide the details of calculation for $\chi^{(1)}_{xx}$, $\chi^{(2)}_{xxx}$, and $\chi^{(3)}_{xxxx}$ in the presence of quenched disorder. As quenched disorder destroys the translational invariance, there is no strong reason for adhering to the periodic boundary condition. We thus choose open boundary condition. A major advantage of open boundary conditions is that it allows one to calculate the linear and nonlinear susceptibilities for other polarizations with only small modifications~\cite{McCoy1971,Derzhko1997}.

The Hamiltonian reads:
\begin{align}
H = -\sum^{L-1}_{n=1}J_n\sigma^z_n \sigma^z_{n+1}-\sum_{n}h_n\sigma^x_n.
\end{align}
Here, $J_n$ and $h_n$ are site dependent, random variables.  We compute all physical observables for several realizations of disorder and then average over these realizations.  

\subsection{Diagonalizing the Hamiltonian}

The diagonalization procedure is similar to the case without disorder. Here, it is more convenient to recast the Jordan-Wigner fermions $c_n$ and $c^\dagger_n$ in terms of even and odd Majorana fermion modes~\cite{Kitaev2001}:
\begin{align}
\alpha_{n} = c^\dagger_n+c^{\phantom{\dagger}}_n;\quad
i\beta_{n} = c^\dagger_n-c^{\phantom{\dagger}}_n.
\end{align}
In particular, $\{\alpha_m,\alpha_n\} = 2\delta_{mn}$, $\{\beta_m,\beta_n\} = 2\delta_{mn}$, and $\{\alpha_m,\beta_n\} = 0$. Furthermore, $\alpha_n$ is even under complex conjugation, whereas $\beta_n$ is odd. The Jordan-Wigner transformed Hamiltonian is then recasted as:
\begin{align}
H &= -i\sum^{L-1}_{n=1}J_n\beta_{n}\alpha_{n+1}-i\sum^{L}_{n=1}h_n\beta_{n} \alpha_{n} \nonumber\\
&= \frac{i}{4}(\alpha^t\mathrm{J}^t\beta-\beta^t\mathrm{J}\alpha).
\end{align}
Here, $\alpha,\beta$ are understood as the $L\times 1$ column vectors formed by $\alpha_i$ and $\beta_i$, respectively. $\mathrm{J}$ is a $L\times L$ bidiagonal matrix:
\begin{align}
\mathrm{J} = \begin{pmatrix}
2h_1 & 2J_1 \\
& 2h_2 & 2J_2\\
& & 2h_3 & 2J_3 \\
& & & \ddots & \ddots \\
& & & & 2h_{L-1} & 2J_{L-1} \\
& & & & & 2h_L
\end{pmatrix}.
\end{align}

We now seek the appropriate orthogonal transformation over $\alpha,\beta$ that brings the above Hamiltonian into diagonal form. To this end, we perform a singular value decomposition (SVD) on $\mathrm{J}$: $\mathrm{J} = \mathrm{U\Lambda V^t}$. Inserting the SVD into the above, we find,
\begin{align}
H &= \frac{i}{4}(\alpha^t\mathrm{V\Lambda}\mathrm{U}^t\beta-\beta^t\mathrm{U\Lambda}\mathrm{V}^t\alpha)\nonumber\\
& = \frac{i}{4}(\gamma^t\mathrm{\Lambda}\delta-\delta^t\mathrm{\Lambda}\gamma) = \frac{i}{4}\sum_n \lambda_n (\gamma_n \delta_n-\delta_n \gamma_n) = i\sum_{n}\lambda_n(d^\dagger_n d^{\phantom\dagger}_n-\frac{1}{2}).
\end{align}
In the second line, we have defined new Majorana modes $\gamma \equiv \mathrm{V}^t\alpha$, and $\delta\equiv\mathrm{U}^t\beta$. They obey a similar set of anti-commutation relations: $\{\gamma_m,\gamma_n\} = 2\delta_{mn}$, $\{\delta_m,\delta_n\} = 2\delta_{mn}$, and $\{\gamma_m,\delta_n\} = 0$. Furthermore, $\gamma_n$ are even under complex conjugation and $\delta_n$ are odd. Now modes carrying different labels $n$ are independent.  In the last equation, we have defined the complex fermions $d_n = (\gamma_n +i\delta_n)/2$. Clearly, $d_n$ annihilates vacuum.

It is useful for latter purpose to define a $2L\times 2L$ correlation matrix $\mathrm{C}(t)$, whose matrix elements are defined as:
\begin{align}
\mathrm{C}_{2m-1,2n-1}(t) &= \langle \alpha_m(t) \alpha_n(0)\rangle,\nonumber\\
\mathrm{C}_{2m,2n}(t) &= \langle i\beta_m(t) i\beta_n(0)\rangle,\nonumber\\
\mathrm{C}_{2m-1,2n}(t) &= \langle \alpha_m(t) i\beta_n(0)\rangle,\nonumber\\
\mathrm{C}_{2m,2n-1}(t) &= \langle i\beta_m(t) \alpha_n(0)\rangle.
\end{align}
To find its explicit value, we calculate a similar correlation matrix $\mathrm{K}(t)$ for $\gamma_n$ and $\delta_n$. The non-zero matrix elements of $\mathrm{K}(t)$ are:
\begin{align}
\mathrm{K}_{2n-1,2n-1}(t) &= \langle \gamma_n(t) \gamma_n(0)\rangle = \cos(\lambda_n t)-i\tanh(\beta\lambda_n/2)\sin(\lambda_n t),\nonumber\\
\mathrm{K}_{2n,2n}(t) &= \langle i\delta_n(t) i\delta_n(0)\rangle = -\cos(\lambda_n t)+i\tanh(\beta\lambda_n/2)\sin(\lambda_n t),\nonumber\\
\mathrm{K}_{2n-1,2n}(t) &= \langle \gamma_n(t) i\delta_n(0)\rangle = i\sin(\lambda_n t)-\tanh(\beta\lambda_n/2)\cos(\lambda_n t),\nonumber\\
\mathrm{K}_{2n,2n-1}(t) &= \langle i\delta_n(t) \gamma_n(0)\rangle = -i\sin(\lambda_n t)+\tanh(\beta\lambda_n/2)\cos(\lambda_n t).
\end{align}
$\mathrm{C}$ is simply related to $\mathrm{K}$ through an orthogonal transformation:
\begin{align}
\mathrm{C}(t) = \begin{pmatrix}
\mathrm{V} & 0 \\
0 & \mathrm{U}
\end{pmatrix}\mathrm{K}(t)\begin{pmatrix}
\mathrm{V}^t & 0\\
0 & \mathrm{U}^t
\end{pmatrix}.
\end{align}

\subsection{Computing the susceptibilities}

We are now ready to calculate the susceptibilities for the polarization along $\hat{x}$ for the disordered case. We begin with the total magnetization:
\begin{align}
M^x_n = \frac{1}{2}\sum^{L-R}_{n=R+1}\sigma^x_n \label{eq:def_mx}.
\end{align}
Note the leftmost $R$ sites and the rightmost $R$ sites of the chain are excluded from the summation. This is to avoid the effects from the zero energy edge modes. The choice of $R$ depends on the correlation length.

To compute the susceptibilities, we first relate them to multi-point spin-spin correlation functions:
\begin{subequations}
\begin{align}
\chi^{(1)}_{xx}(t) &= \frac{i\theta(t)}{L}\langle [M^x(t), M^x(0)] \rangle  = \frac{\theta(t)}{2L}\sum_{j,l}\mathrm{Im} S^x_{jl}(t,0).
\end{align}
Here, the first equality follows from the Kubo formula. The second equality follows from expanding the commutator. $S^x_{jl}(s_1,s_2) = \langle \sigma^x_{j}(s_1)\sigma^x_{l}(s_2)\rangle$ is the two-point spin correlation function in vacuum. Likewise,
\begin{align}
\chi^{(2)}_{xxx}(t,\tau+t) &= \frac{i^2 \theta(t)\theta(\tau)}{L}\langle [[M^x(\tau+t), M^x(\tau)], M^x(0)] \rangle \nonumber\\
& = -\frac{\theta(t)\theta(\tau)}{4L}\sum_{j,l,m}\mathrm{Re} S^x_{jlm}(\tau+t,t,0) - \mathrm{Re} S^x_{jlm}(\tau,\tau+t,0),
\end{align}
where $S^x_{jlm}(s_1,s_2,s_3) = \langle \sigma^x_j(s_1) \sigma^x_l(s_2) \sigma^x_m (s_3)\rangle$ is the three-point spin correlation function. 
\begin{align}
\chi^{(3)}_{xxxx}(t_3,t_2+t_3,t_1+t_2+t_3) = \frac{i^3 \theta(t_1)\theta(t_2)\theta(t_3)}{L}\langle [[[M^x(t_1+t_2+t_3), M^x(t_1+t_2)], M^x(t_1)], M^x(0)] \rangle \nonumber\\
 = \frac{\theta(t_1)\theta(t_2)\theta(t_3)}{8L}\sum_{j,l,m,n}\mathrm{Im}S^x_{jlmn}(t_1+t_2+t_3,t_1+t_2,t_1,0)+\mathrm{Im}S^x_{jlmn}(t_1,t_1+t_2,t_1+t_2+t_3,0) \nonumber\\
+\mathrm{Im}S^x_{jlmn}(0,t_1+t_2,t_1+t_2+t_3,t_1)+\mathrm{Im}S^x_{jlmn}(0,t_1,t_1+t_2+t_3,t_1+t_2),
\end{align}
\end{subequations}
where $S^x_{jlmn}(s_1,s_2,s_3,s_4) = \langle \sigma^x_j(s_1) \sigma^x_l(s_2) \sigma^x_m (s_3) \sigma^x_n (s_4) \rangle$ is the four-point spin correlation function. 

Thus the task of computing susceptibilities is reduced to computing spin correlation functions. To this end, we express $\sigma^x_n$ in terms of Majorana modes $\alpha_n$ and $\beta_n$, and then compute the average using the Wick theorem. The results are compactly expressed as matrix Pfaffians~\cite{McCoy1971}.

The two-point spin correlation function:
\begin{subequations}
\begin{align}
S^x_{jl}(s_1,s_2) = \mathrm{Pf}\begin{pmatrix}
\mathrm{C}_{2j-1:2j,2j-1:2j}(0) & \mathrm{C}_{2j-1:2j,2l-1:2l}(s_1-s_2) \\
 & \mathrm{C}_{2l-1:2l,2l-1:2l}(0)
\end{pmatrix}.
\end{align}  
Here, $\mathrm{C}_{2j-1:2j,2l-1:2l}(t)$ stands for the $2\times 2$ sub-matrix of the previously defined correlation matrix $\mathrm{C}(t)$. The first index runs from $2j-1$ to $2j$, and the second runs from $2l-1$ to $2l$. The other entities are similarly defined. Since the argument of a Pfaffian must be a skew-symmetric matrix, it is understood that diagonal entries have all been set to 0. The other spin correlation functions are found in the same vein:
\begin{align}
S^x_{jlm}(s_1,s_2,s_3) = -\mathrm{Pf}\begin{pmatrix}
\mathrm{C}_{2j-1:2j,2j-1:2j}(0) & \mathrm{C}_{2j-1:2j,2l-1:2l}(s_1-s_2) & \mathrm{C}_{2j-1:2j,2m-1:2m}(s_1-s_3) \\
{} &  \mathrm{C}_{2l-1:2l,2l-1:2l}(0) & \mathrm{C}_{2l-1:2l,2m-1:2m}(s_2-s_3)\\
{} & {} & \mathrm{C}_{2m-1:2m,2m-1:2m}(0) 
\end{pmatrix},
\end{align}
and
\begin{align}
& S^x_{jlmn}(s_1,s_2,s_3,s_4) \nonumber\\
& = \mathrm{Pf}\begin{pmatrix}
\mathrm{C}_{2j-1:2j,2j-1:2j}(0) & \mathrm{C}_{2j-1:2j,2l-1:2l}(s_1-s_2) & \mathrm{C}_{2j-1:2j,2m-1:2m}(s_1-s_3) & \mathrm{C}_{2j-1:2j,2n-1:2n}(s_1-s_4) \\
{} & \mathrm{C}_{2l-1:2l,2l-1:2l}(0) & \mathrm{C}_{2l-1:2l,2m-1:2m}(s_2-s_3) & \mathrm{C}_{2l-1:2l,2n-1:2n}(s_2-s_4) \\
{} & {} & \mathrm{C}_{2m-1:2m,2m-1:2m}(0) & \mathrm{C}_{2m-1:2m,2n-1:2n}(s_3-s_4) \\
{} & {} & {} & \mathrm{C}_{2n-1:2n,2n-1:2n}(0)
\end{pmatrix}.
\end{align}
\end{subequations}

In our calculation, we use a a chain of 40 sites, and set $R=5$ (Eq.~\ref{eq:def_mx}). We first compute the susceptibilities in real time and then perform a numerical FFT. The choice of time grid is identical to Sec.~\ref{sec:fft}. We compute the matrix Pfaffian by using the numerical package developed by Wimmer~\cite{Wimmer2012}.

\section{Calculating the 2D spectra along $\hat{y}$}

\subsection{Computing the susceptibilities}

Here we provide the details for calculating the susceptibilities with polarization along $\hat{y}$. The procedure is largely in parallel with Section~\ref{sec:chi_x_disorder}: we first relate the susceptibilities to multi-point spin correlation functions, and then compute the spin correlation functions by going to the Jordan-Wigner fermion basis. Since this section concerns the susceptibilities along $\hat{y}$, we need to find the multi-point spin correlation functions $S^y_{jl}(s_1,s_2)$ and $S^y_{jlmn}(s_1,s_2,s_3,s_4)$. Note the second-order susceptibility $\chi^{(2)}_{yyy}$ vanishes by symmetry.

\begin{figure}
\includegraphics[width = 0.8\textwidth]{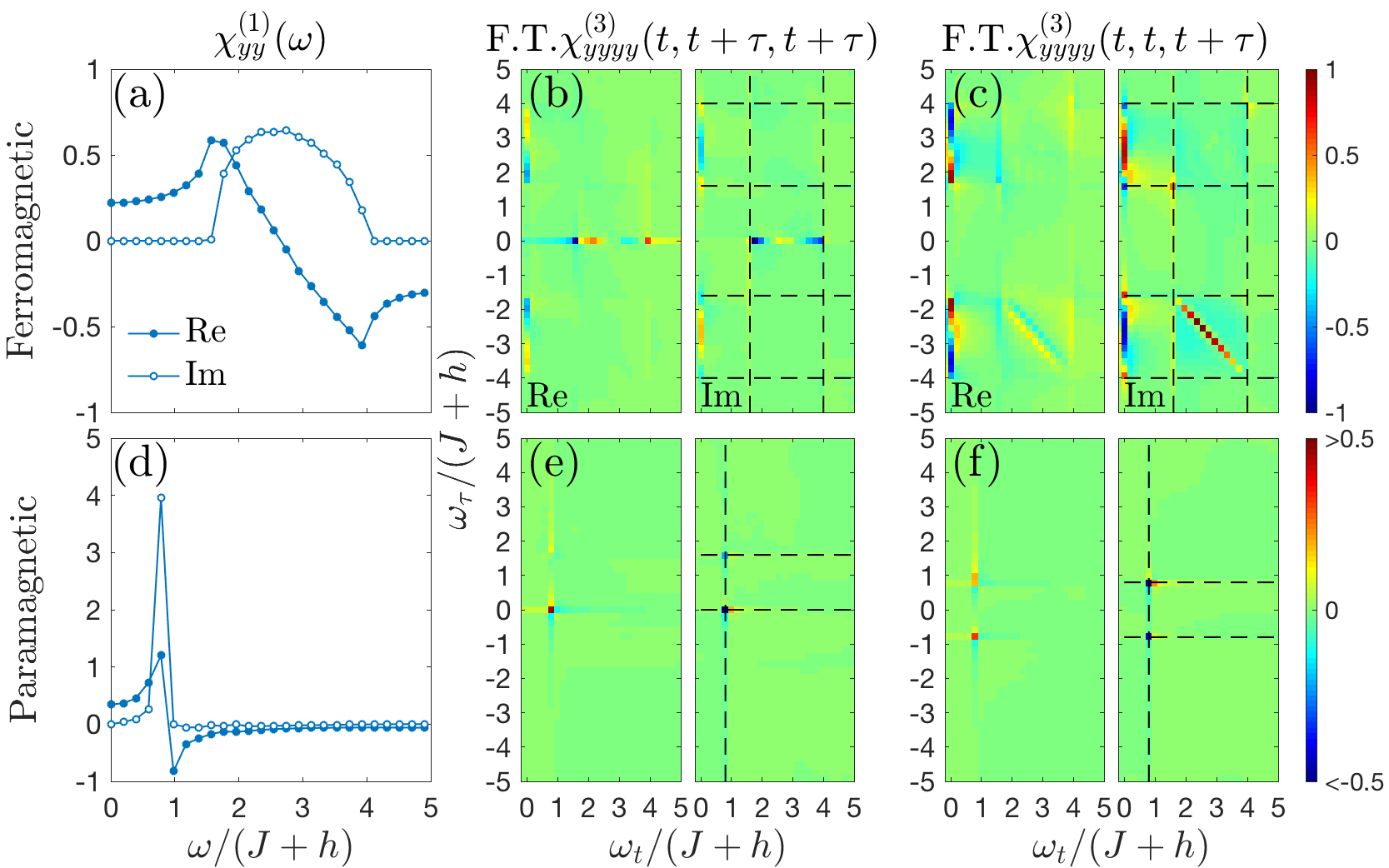}
\caption{1D and 2D spectra for polarization along $\hat{y}$. The top row is for the ferromagnetic phase ($h/(h+J)=0.3$), whereas the bottom row is for the paramagnetic phase ($h/(h+J)=0.7$). For the 2D spectra, only the first and fourth quadrant are shown. The other half are obtained by complex conjugation. The data are rescaled such that the maximal absolute value of the Fourier transform is 1. Note the 2D spectra in the bottom row is color saturated to reveal weaker features. The dashed lines in (b) and (c) demarcate the boundaries, $[1.6(J+h),4(J+h)]$, of the spinon continuum. The dashed lines in (e) and (f) mark on the frequency axes the locations of $0$, $\pm\omega_{k=0}$, and $2\omega_{k=0}$, where $\omega_{k=0} = 0.8 (J+h)$ is the fundamental frequency of the optical magnon in the paramagnetic phase.}
\label{fig:chi_y}
\end{figure}

The spin operator $\sigma^y_n$ is non-local in the Jordan-Wigner fermion basis:
\begin{align}
\sigma^y_n = \alpha_1 i\beta_2 \alpha_2 i\beta_2 \cdots \alpha_{n-1} i\beta_{n-1} i\beta_{n}.
\end{align}
Thus, the two-point spin correlation function becomes a highly non-local correlation function in the fermion basis. Nevertheless, it may be compactly expressed as matrix Pfaffian~\cite{Derzhko1997}:
\begin{subequations}
\begin{align}
S^y_{jl}(s_1,s_2) = \mathrm{Pf}\begin{pmatrix}
\mathrm{C}_{1:2j-2:2j,1:2j-1:2j}(0) & \mathrm{C}_{1:2j-2:2j,1:2l-1:2l}(s_1-s_2) 
\\
{} & \mathrm{C}_{1:2l-2:2l,1:2l-2:2l}(0)
\end{pmatrix}.
\end{align}
Here, $\mathrm{C}_{1:2j-2:2j,1:2l-2:2l}(t)$ stands for the $(2j-1) \times (2l-1)$ sub-matrix of the aforementioned correlation matrix $\mathrm{C}(t)$. The first index is taken from the set, $\{1,2,3,\cdots 2j-2, 2j\}$, and the second from $\{1,2,3,\cdots 2l-2, 2l\}$. The other objects are similarly defined. The four-point correlation function is given in a similar form:
\begin{align}
& S^y_{jlmn}(s_1,s_2,s_3,s_4) \\
& = \mathrm{Pf}\begin{pmatrix}
\mathrm{C}_{1:2j-2:2j,1:2j-2:2j}(0) & \mathrm{C}_{1:2j-2:2j,1:2l-2:2l}(s_1-s_2) & \mathrm{C}_{1:2j-2:2j,1:2m-2:2m}(s_1-s_3) & \mathrm{C}_{1:2j-2:2j,1:2n-2:2n}(s_1-s_4) \\
{} & \mathrm{C}_{1:2l-2:2l,1:2l-2:2l}(0) & \mathrm{C}_{1:2l-2:2l,1:2m-2:2m}(s_2-s_3) & \mathrm{C}_{1:2l-2:2l,1:2n-2:2n}(s_2-s_4) \\
{} & {} & \mathrm{C}_{1:2m-2:2m,1:2m-2:2m}(0) & \mathrm{C}_{1:2m-2:2m,1:2n-2:2n}(s_3-s_4) \\
{} & {} & {} & \mathrm{C}_{1:2n-2:2n,1:2n-2:2n}(0) \nonumber\\
\end{pmatrix}.
\end{align}
\end{subequations}
Apparently, computing $S^y_{jlmn}(s_1,s_2,s_3,s_4)$ is considerably more expensive comparing to $S^x_{jlmn}(s_1,s_2,s_3,s_4)$ due to the large size of the matrix.  We benchmark this procedure on a short chain with $L=8$. We compute the susceptibilities by using two methods: (a) a direct numerical diagonalization of the spin Hamiltonian followed by the Kubo formula, and (b) the Pfaffian-based method. The absolute error is within $10^{-13}$ for all cases considered.

\subsection{Results for ferromagnetic and paramagnetic phases}

In this section, we present the 2D spectra in the ferromagnetic phase and the paramagnetic phase for the $y$ polarization. The calculation set up is identical to Sec.~\ref{sec:chi_x_disorder}. We shall focus on the case without disorder or spinon decay.  We first consider the ferromagnetic phase. We set the model parameter to the representative value $h/(h+J) = 0.3$. As the operator $\sigma^y_n$ flips the eigenvalues of $\sigma^z_{n}$ and thereby creating domain walls, we expect the spectra along $\hat{y}$ is qualitatively similar to $\hat{x}$. The calculation confirms this picture. The imaginary part of the linear susceptibility, $\mathrm{Im}\chi^{(1)}_{yy}$, shows a spinon continuum virtually identical to $\mathrm{Im}\chi^{(1)}_{xx}$ (Fig.~\ref{fig:chi_y}(a)). The 2D spectra  (Fig.~\ref{fig:chi_y}(b)(c)) are also qualitatively similar to the spectra along $\hat{x}$. In particular, we observe a strong rephasing streak in the fourth quadrant of Fig.~\ref{fig:chi_y}c. The streak runs along the diagonal direction, mirroring the energy range of the spinon pairs, whereas its width along the anti-diagonal direction is resolution limited, reflecting the infinite life time of the spinon pairs. The weak, box-shaped shadow is likely due to the fact that the $\sigma^y_n$ operator is non local in the spinon basis.

We next turn to the paramagnetic phase, where the spins are ordered in $\sigma^z_n$. We set the model parameters to the representative value $h/(J+h) = 0.7$. For the pulse polarization along $\hat{y}$, the optically active excitation is the magnon with momentum $k=0$, and its energy $\omega_{k=0} = 2(h-J) = 0.8(J+h)$.  We observe a sharp resonance at $\omega_{k=0}$ in $\mathrm{Im}\chi^{(1)}_{yy}$ consistent with the lack of fractionalized excitation in the paramagnetic regime.  In the 2D spectra, the various peaks and their locations are expected for nonlinear spin waves~\cite{lu2017coherent}. In the Fourier transform of $\chi^{(3)}_{yyyy}(t,\tau+t,\tau+t)$, we observe a strong peak at the pumping frequency $\omega_\tau = 0$ and the detection frequency $\omega_t = \omega_{k=0}$, known as the pump-probe signal, and a weaker peak at the pumping frequency $\omega_\tau = 2\omega_k$ and the detection frequency $\omega_t = \omega_{k=0}$, known as the two-quanta signal~\cite{lu2017coherent}. On the other hand, the Fourier transform of $\chi^{(3)}_{yyyy}(t,t,\tau+t)$ contains a non-rephasing peak at $\omega_\tau = \omega_{k=0}$ and $\omega_t = \omega_{k=0}$, and a rephasing peak at $\omega_\tau = -\omega_{k=0}$ and $\omega_t = \omega_{k=0}$.

\section{Diagonal transitions and frequency vectors}

 Two-level systems are paradigmatic models for nonlinear optics~\cite{shen1984principles}. In a standard treatment such as in Ref.~\cite{shen1984principles}, it is assumed that the light can induce transitions between the ground state and the excited state. The ``diagonal'' transitions. i.e. from the ground state to the ground state, or from the excited state to the excited state, are usually not included in the model as they are often forbidden by symmetry. In this work however, the two-level systems that emerge from the transverse field Ising chain do not represent atomic orbitals but rather the spinon pairs.  Diagonal transitions are allowed in this system. These diagonal transitions give rise to novel features in the 2D coherent spectra including a non-rephasing-\emph{like} (NR-like) signal in $\chi^{(2)}$, and a terahertz-rectification-\emph{like} (TR-like) signal in $\chi^{(3)}$, which should be absent in the 2D spectra of a canonical two-level system. These novel features may appear unusual at first glance. For instance, the TR is often thought of as a $\chi^{(2)}$ process. In this work, however, their signals seems to appear in the spectra associated with $\chi^{(3)}$ as well. As we shall see, the TR-\emph{like} signals in $\chi^{(3)}$ spectra do not reflect true nonlinear rectification processes, but instead result from the aforementioned diagonal transitions. In other words, these TR-\emph{like} signals are not true TR signals but merely resemble them in the frequency plane.

In this section, we elucidate the impact of the diagonal transitions on 2D coherent spectra by analyzing a toy model. We will also connect our results to the ``frequency vector'' scheme for understanding the 2D spectra~\cite{[{}][. {Note the positions of the frequency vectors used therein differ from ours due to different parameterization of the time variables.}] kuehn2011nonlinear}. The purpose of this section is pedagogical. It introduces the frequency vector scheme to readers that are unfamiliar with it, while, for readers with a background in nonlinear optics, it aims at showing how the aforementioned novel features in the 2D spectra can be understood by using a familiar language.

We first consider a standard two-level system.  Written in spin language, the Hamiltonian is:
\begin{align}
H = \frac{\Omega}{2}\sigma^z.
\end{align}
Here, $\sigma^z = -1$ ($\sigma^z = 1$) in the ground (excited) state.  $\Omega>0$ is the energy level splitting.

Suppose we perform a THz 2D coherent spectroscopy experiment on this system. We use two linearly polarized, Dirac-$\delta$ pulses. The magnetic field is given by:
\begin{align}
B^\alpha(s) = \epsilon^\alpha[A_0 \delta(s)+A_\tau\delta(s-\tau)].
\end{align}
Here, $s$ is the time, $A_{0,\tau}$ the pulse area, $\epsilon^\alpha$ the polarization vector, which we set to be:
\begin{align}
\epsilon^x = 0;\quad
\epsilon^y = \sin\theta;\quad
\epsilon^z = \cos\theta. 
\end{align}
Here, $\theta$ is the angle between the polarization vector and the $\hat{z}$ axis. We measure the response parallel to $\epsilon^\alpha$. As our purpose is pedagogical, we will omit dissipation for simplicity.

Since the $\hat{z}$ component of the magnetic field pulse is non-vanishing when $\theta\neq \pi/2$, diagonal transitions are allowed, and consequently the 2D spectra should contain the novel features mentioned in the beginning of this section. Alternatively, we may think that mixed nonlinear response functions such as $\chi^{zyy}$ contribute to the 2D spectra. These two points of view are equally valid and complementary to each other. In what follows, we compute the 2D spectra of this system and relate these results to the frequency vector scheme.

\subsection{Contributions to $\chi^{(2)}$}

We first consider the $\chi^{(2)}$. To this end, we switch to the Heisenberg picture: $\sigma^y (t) = \cos(\Omega t)\sigma^y+\sin(\Omega t)\sigma^x$, and $\sigma^z(t) = \sigma^z$. The latter follows from the fact that $\sigma^z$ commutes with $H$. A straightforward calculation using the Kubo formula shows:
\begin{subequations}
\begin{align}
\chi^{(2)} = \sin^2\theta \cos\theta (\chi^{(2)}_{zyy}+\chi^{(2)}_{yzy}).
\end{align}
Here, the trigonometric pre-factor is due to the polarization vector. The two mixed response functions are given by:
\begin{align}
\chi^{(2)}_{zyy}(t,t+\tau) &= 4\theta(t)\theta(\tau)\cos(\Omega \tau);  \\
\chi^{(2)}_{yzy}(t,t+\tau) &= -4\theta(t)\theta(\tau)\cos[\Omega(t+\tau)].
\end{align}
\end{subequations}

We observe that $\chi^{(2)} = 0$ when the diagonal transition is absent, i.e. $\theta = \pi/2$. $\chi^{(2)}_{zyy}$ is a function of $\tau$ only. In the frequency domain, this gives rise to a peak at $\omega_\tau = \Omega $, and $\omega_t= 0$. This is the THz rectification (TR) signal. However, $\chi^{(2)}_{yzy}$ gives rise to a non-rephasing-like (NR-like) signal at $\omega_\tau = \omega_t = \Omega$. This is remarkable because usually NR signals occur in $\chi^{(3)}$.

\begin{figure}
\includegraphics[width = 0.4\textwidth]{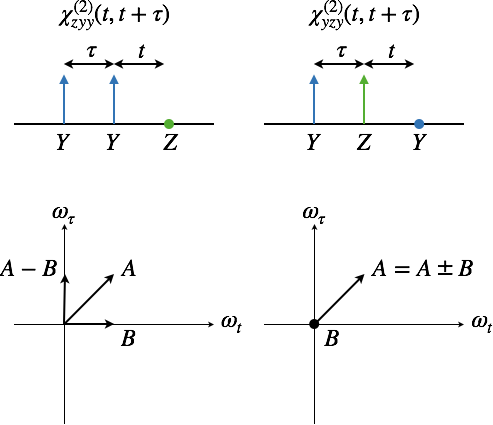}
\caption{The two $\chi^{(2)}$ response functions that contribute to the 2D spectra. The top row shows the pulse sequences, where the arrows represent the Dirac-$\delta$ pulses and the dot represents the sample response. Latin scripts denote the polarization. Bottom row shows the corresponding frequency vectors. Each pulse is mapped to a vector in the frequency plane spanned by $\omega_t$ and $\omega_\tau$. The nonlinear signals are then understood as linear combinations of these frequency vectors.}
\label{fig:chi2_freq_vec}
\end{figure}

We may rationalize these results by using a scheme for 2DCS known as frequency vectors. One may view it as a generalization of the usual frequency sum/difference picture discussed in nonlinear optics~\cite{shen1984principles}. First consider $\chi^{(2)}_{zyy}$ (Fig.~\ref{fig:chi2_freq_vec}, left). For this response function, both pulses are polarized along $\hat{y}$. Since $\sigma^y$ induces transitions between the ground state and the excited state, the first pulse will induce an oscillatory signal of the form $\exp[i\Omega(t+\tau)]$ and the second an oscillatory signal of the form $\exp(i\Omega t)$. These two oscillations are represented by two non-orthogonal vectors in the frequency plane spanned by $\omega_\tau$ and $\omega_t$~\cite{kuehn2011nonlinear}: the first pulse (pulse A) gives a diagonal vector, whereas the second pulse (pulse B) gives a horizontal vector. The $\chi^{(2)}$ nonlinear coupling produce a difference vector (A$-$B) along the $\omega_\tau$ axis. This is precisely the TR signal we found by calculation. The sum vector (A$+$B) would produce a second-harmonic generation signal. However, it is absent in the two-level system as such a system can only emit radiation at fixed frequency $\Omega$. Note that we only consider the signals in the first and fourth quadrant of the frequency plane. Signals in the other two quadrants are related by reflection.

Somewhat different considerations apply to $\chi^{(2)}_{yzy}$ (Fig.~\ref{fig:chi2_freq_vec}). In this case, the first pulse (A) is polarized along $\hat{y}$, which gives a diagonal vector (A) in the frequency plane. Importantly, the second pulse (B) is polarized along $\hat{z}$, and such a pulse does not induce any oscillations. We may represent this as a null vector (B) in the frequency plane. The sum or difference of the two vectors (A$\pm$B) are thus the NR-like signal we found by calculation.

\subsection{Contributions to $\chi^{(3)}$}

Having understood $\chi^{(2)}$, we can now move to $\chi^{(3)}$. A straightforward calculation yields,
\begin{subequations}
\begin{align}
\chi^{(3)} = \sin^4\theta\chi^{(3)}_{yyyy}+\sin^2\theta\cos^2\theta (\chi^{(3)}_{zyzy}+\chi^{(3)}_{yzzy}),
\end{align}
where
\begin{align}
\chi^{(3)}_{yyyy}(t_3,t_2+t_3,t_1+t_2+t_3) &= -8\theta(t_1)\theta(t_2)\theta(t_3)\sin(\Omega t_3)\cos(\Omega t_1); \\
\chi^{(3)}_{zyzy}(t_3,t_2+t_3,t_1+t_2+t_3) &= 8\theta(t_1)\theta(t_2)\theta(t_3)\sin[\Omega(t_1+t_2)]; \\
\chi^{(3)}_{yzzy}(t_3,t_2+t_3,t_1+t_2+t_3) &= -8\theta(t_1)\theta(t_2)\theta(t_3)\sin[\Omega(t_1+t_2+t_3)].
\end{align}
\end{subequations}

\begin{figure}
\includegraphics[width = 0.6\textwidth]{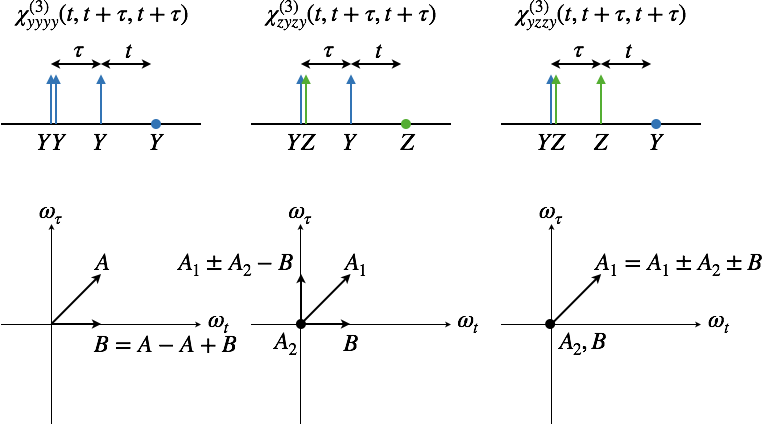}
\caption{The three nonlinear response functions that contribute to $\chi^{(3)}(t,t+\tau,t+\tau)$ and the corresponding frequency vectors.}
\label{fig:chi3_aab_freq_vec}
\end{figure}

The experimentally measured $\chi^{(3)}$'s are the two special limits of $\chi^{(3)}(t_3,t_2+t_3,t_1+t_2+t_3)$. $\chi^{(3)}(t,t+\tau,t+\tau)$ corresponding to the limit $t_1\to 0, t_2\to\tau, t_3\to t$. The nonlinear response $\chi^{(3)}_{yyyy}(t,t+\tau,t+\tau)$ produces a pump-probe (PP) signal at $\omega_\tau = 0$ and $\omega_t = \Omega$. This response function corresponds to a pulse sequence in which all three pulses are polarized along $\hat{y}$ (Fig.~\ref{fig:chi3_aab_freq_vec}, left). In terms of the frequency vectors, this means the first two pulses gives a diagonal vector (A), and the third pulse gives a horizontal vector (B). The PP signal thus comes from the linear combination A$-$A$+$B. The other combinations, namely  A$+$A$\pm$B, while in principle possible, are absent for the two-level system.  

$\chi^{(3)}_{zyzy}(t,t+\tau,t+\tau)$ produces a TR-like signal at $\omega_\tau = \Omega$ and $\omega_t = 0$. This response corresponds to a pulse sequence in which the first pulse is polarized along $\hat{y}$, the second along $\hat{z}$, and the third along $\hat{y}$ (Fig.~\ref{fig:chi3_aab_freq_vec}, middle). In terms of frequency vectors, the first gives a diagonal vector (A$_1$), the second a null vector (A$_2$), and the third a horizontal vector (B). The TR-like signal comes from the combination A$_1\pm$A$_2-$B. The other combination A$_1\pm$A$_2+$B would give rise to a second-harmonic generation signal, which is absent for the two-level system.

$\chi^{(3)}_{yzzy}(t,t+\tau,t+\tau)$ produces a NR signal at $\omega_\tau = \omega_t = \Omega$. The pulse polarizations are successively along $\hat{y}$, $\hat{z}$, and $\hat{z}$ (Fig.~\ref{fig:chi3_aab_freq_vec}, right). In the frequency plane, the first pulse gives a diagonal vector (A$_1$), and the second and the last null vectors (A$_2$ and B). Their combinations (A$_1\pm$A$_2\pm$B) has a unique outcome, namely the NR signal.

\begin{figure}
\includegraphics[width = 0.6\textwidth]{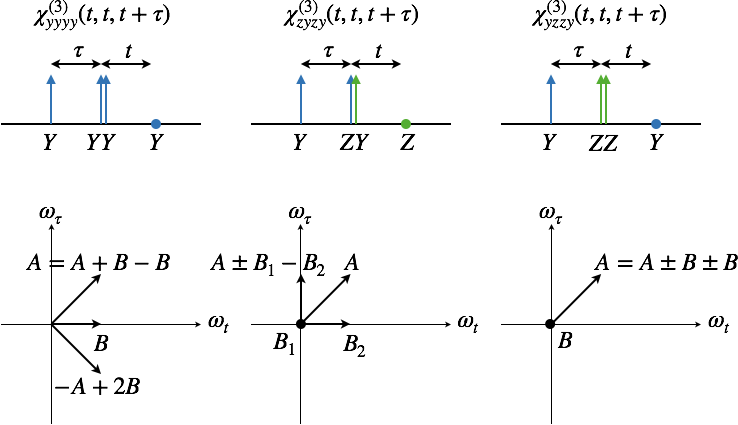}
\caption{The same as Fig.~\ref{fig:chi3_aab_freq_vec} but for $\chi^{(3)}(t,t,t+\tau)$.}
\label{fig:chi3_abb_freq_vec}
\end{figure}

$\chi^{(3)}(t,t,t+\tau)$ correspond to the limit $t_1\to \tau, t_2\to0, t_3\to t$. $\chi^{(3)}_{yyyy}(t,t,t+\tau)$ produces a NR signal and a rephasing signal; $\chi^{(3)}_{zyzy}(t,t,t+\tau)$, a TR-like signal; $\chi^{(3)}_{yzzy}(t,t,t+\tau)$, a NR signal. These signals can also be understood in terms of frequency vectors as shown in Fig.~\ref{fig:chi3_abb_freq_vec}. For $\chi^{(3)}_{yyyy}(t,t,t+\tau)$, the first pulse (A) gives a diagonal vector, and the second and third a horizontal vector (B). The combination A$+$B$-$B yields the NR signal, whereas the other combination $-$A$+2$B gives the rephasing signal. For $\chi^{(3)}_{zyzy}(t,t,t+\tau)$, the first pulse (A) gives a diagonal vector, the second a null vector (B$_1$), and the third a horizontal vector (B$_1$). Their combinations A$\pm$B$_1-$B$_2$ yield the TR-like signal. Finally, for $\chi^{(3)}_{yzzy}(t,t,t+\tau)$, the first pulse gives a diagonal vector (A), and the second and third null vectors (B). Their combinations A$\pm$B$_1\pm$B$_2$ yield the NR signal.

\end{document}